\documentclass[prd,preprint,a4paper,showkeys,showpacs]{revtex4-1} 

\usepackage{bm}
\usepackage{amsmath}
\usepackage{amssymb}
\usepackage{graphicx}
\usepackage{color}

\def\p{\partial}

\def\b{\beta}
\def\d{\delta}

\def\ve{\varepsilon}
\def\k{\kappa}
\def\l{\lambda}
\def\L{\Lambda}

\def\f{\phi}

\def\th{\theta}

\def\ra{\rightarrow}

\def\Red{}

\begin{document}

\title{Renormalization group for effective field theories:\\ 
cutoff schemes and universality}

\author{Jos\'e Gaite}
\affiliation{
Applied Physics Dept.,
ETSIAE,
Universidad Polit\'ecnica de Madrid,\\ E-28040 Madrid, Spain}

\date{September 1, 2025}

\begin{abstract}
In effective field theories, the concept of renormalization of perturbative divergences 
is replaced by renormalization group concepts such as 
{\em relevance} and {\em universality}. Universality is related to cutoff scheme independence 
in renormalization. Three-dimensional scalar field theory with just the quartic coupling 
is universal but the less relevant sextic coupling introduces a cutoff scheme dependence, 
which we quantify by three independent parameters, in the two-loop order of 
perturbation theory. However, reasonable schemes only allow reduced 
ranges of those parameters, even contrasting the sharp cutoff with very smooth cutoffs. 
The sharp cutoff performs better. 
In any case, the effective field theory possesses some degree of universality even in 
the massive case (off criticality). 
\end{abstract}


\maketitle

\section{Introduction}
\label{intro}

Theories of the physics 
of systems with a large or infinite number of degrees of freedom must address the problem of the change in the relevant description as the scale of the observation changes. In the old methods of quantum field theory, the problem of the appearance 
of infinite quantities in perturbative calculations gave rise to the theory of renormalization, in which the infinite quantities are assumed to belong to an unobservable domain, while the observable renormalized quantities are always finite \cite{Parisi,ZJ}. 
Renormalized field theory has reached a high degree of sophistication, in particular, with the development of dimensional regularization and the epsilon expansion \cite{ZJ}. 
Within dimensional regularization, infinite quantities take a different aspect, given that there is no high-energy or small-distance cutoffs to consider, and hence the subtleties of taking the infinite cutoff limit do not arise. 
Such methods are best suited for the physics that is independent of what happens at high energy or small distance, as occurs, for example, in the calculation of critical exponents 
in critical phenomena \cite{Parisi,ZJ}. In the domain where scale invariance 
holds, the problem of dealing with very many degrees of freedom is somehow 
compensated for by the scale symmetry, which usually becomes full conformal symmetry. 

At any rate, there are situations in which scale invariance is not exact or in which 
the short-distance physics does not fully decouple from the scales one is interested in. Nevertheless, the physics in the latter scales can be rather different, because renormalization still plays a substantial role. During many years, a systematic set of methods has been developed to study the physics of systems in which the high-energy or small-distance degrees of freedom play a lesser but non-negligible role. These methods are 
the methods of {\em effective field theory} \cite{Cao-S,Burgess}. In effective field theory, 
the cutoff is held finite and one is not concerned with the divergences that arise in the infinite cutoff limit. In fact, the methods of effective field theory greatly overlap with 
the non-perturbative approach to the renormalization group pioneered by Wilson 
\cite{Wil-Kog}, in which renormalization is understood as the effect of 
the progressive removal of short-distance degrees of freedom 
(a procedure called {\em exact renormalization group}, 
or ERG). The renormalized or effective field theory accounts for
the effect of those degrees of freedom through a small number of {\em relevant} 
coupling constants. The theory of a scalar field with only relevant couplings is considered 
{\em universal}, as it is determined by symmetry and space dimension only. 

Just as in classic renormalization methods one can choose different methods to obtain 
finite results, i.e., various {\em regularization} methods, there are also different methods 
to remove short-distance degrees of freedom in the Wilson renormalization group (or exact RG).
The simplest type of cutoff regularization, with a sharp wave-number cutoff, 
corresponds to the wave-number ``slice'' integration in the exact RG 
\cite{Wegner-H}. 
However, this simple procedure is not necessarily optimal and there are reasons to try 
instead {\em smooth} cutoffs \cite{Wil-Kog}. The comparison of various regularization 
methods is gaining interest \cite{Yang-Ruan,Ivan,Gold,Kharuk}. 
Goldstein \cite{Gold} considers the most general smooth wave-number cutoff 
for scalar field theory at one loop order 
in the epsilon expansion and shows that the RG flow is altered but 
the Wilson-Fisher critical exponents are not, in accord with universality expectations.
Of course, it is necessary to consider higher loop orders, as is being done 
in related studies \cite{Yang-Ruan,Ivan,Kharuk}. 

The scheme dependence of the exact RG was a cause for concern years ago \cite{Ball,Liao}. The study by Liao et al on smooth cutoff schemes for the ERG 
\cite{Liao}
concluded that smoother cutoffs tend to minimize the strength of neglected irrelevant  
operators. In other words, optimizing the renormalization group would favor smooth cutoff schemes. It should be noted that the purpose of such optimization was to find the best method for calculating critical exponents \cite{Ball,Liao}, which is not our goal here. 
We have recently pointed out the importance of comparing ERG cutoff schemes off-criticality 
and we have performed a comparison between some standard schemes, finding the sharp cutoff to 
be adequate \cite{II}.

In perturbation theory, since the one-loop order is rather trivial, 
it is worthwhile to develop higher loop orders, using 
a cutoff that is as general as possible. In particular, we consider here how 
Liao et al's suite of schemes \cite{Liao} performs in perturbative scalar field theory
at the two-loop order. 
It should be interesting to see how the results depend on epsilon 
in the epsilon expansion but we encounter considerable difficulties in calculations 
that keep the space dimension arbitrary. Therefore, we prefer to work in a fixed 
space dimension. Rather than being based on four dimensions, as is usual 
\cite{Yang-Ruan,Ivan,Gold}, we find it more interesting to work in 
three dimensions, which has more applications in statistical and condensed matter physics. 
In addition, we employ the background-field method, 
which reduces the number of independent calculations \cite{Yang-Ruan,Ivan,Kharuk}.

We shall consider, in three dimensions, not just the ordinary 
$\lambda\phi^4$ theory but the full $(\lambda\phi^4+g\phi^6)_3$ theory, which 
contains all the power-counting renormalizable couplings. Of course, the effect of 
the non-relevant but only {\em marginal} coupling $g\phi^6$ is expected to be small, 
but the effect of changes of cutoff scheme is also expected to be small, 
provided we consider the domain with approximate scale invariance. 
In fact, we will show that both types of effects are connected. 

For completeness, we could also consider, on the one hand, 
the effect of fully irrelevant couplings in the effective potential, and,
on the other, the effect of field renormalization, 
related to the other marginal term, $(\p\phi)^2$. 
Regarding the former, 
fully irrelevant coupling constants depend on the wave-number cutoff as 
inverse powers and become less important at 
low energies or long distances, 
and regarding the latter, its effect is small in $d=3$. 
We shall comment on these matters as they arise.


The text consists of five main sections and a concluding section. 
There are also two appendices, the first one containing the calculation of 
three necessary integrals with different cutoffs, 
and the second one the calculation of beta-functions. 
Section \ref{phi4-phi6} lays the groundwork for the study of scalar field theory in 
three dimensions, comparing the roles of the quartic and sextic couplings. 
Section \ref{phi4} summarizes standard results of three-dimensional $\lambda\phi^4$ theory
and introduces the use of ERG integrations, initiated in Ref.~\onlinecite{II} 
and to be further developed in the following sections.
Section \ref{2loop} introduces the two-loop effective potential and carries out 
the renormalization of the quartic and sextic coupling constants, which 
brings about a cutoff dependence. 
Section \ref{non-univ} studies the non-universal aspects due to 
the sextic coupling, first with a sharp-cutoff example, and then carefully 
studying the modifications produced by smooth cutoffs. 
Section \ref{improved} improves the two-loop renormalization formulas by means 
of the RG that is derived from them. The consequent RG flow provides a fuller picture 
of both renormalization and cutoff dependence.
This picture is explained in the concluding section.

\section{Renormalization of scalar field theory in three-dimensional space}
\label{phi4-phi6}

The renormalization of scalar field theory is suitable for a simple 
comparison of various regularization methods \cite{Yang-Ruan,Ivan,Gold}.
Furthermore, scalar field theory in three-dimensional space, with one or several fields, 
has many applications in statistical and condensed matter physics \cite{Parisi,ZJ,Fradkin}.

In fact, the theory with a scalar field $\f$ with $\f^4$ self-interaction 
is thoroughly studied as the universality class of the Ising model, a paradigm 
in statistical physics for the theory of critical phenomena \cite{Parisi,ZJ}.
This model illustrates the phase transition physics of a simple critical point, but 
there are more complex phase transitions. 
For example, the phase transitions near a tricritical point are quite convoluted 
\cite{LL,Pfeu-T,Law-Sar}. 
Previously, the theory of a scalar field $\f$ with $\f^4$ and $\f^6$ self-interaction
in three dimensions was studied to describe tricritical phenomena 
\cite{Law-Sar,Stephen,Carvalho,Soko}. 
We shall consider this field theory, denoted as $(\lambda\phi^4+g\phi^6)_3$, 
as our model scalar field theory in three dimensions, 
since it 
contains all the power-counting renormalizable couplings. Of course, the effect of 
the non-relevant but just {\em marginal} coupling $g\phi^6$ is expected to be small 
\cite{Stephen}, but it deserves further study.

Some years ago, 
the renormalization of $(\lambda\phi^4+g\phi^6)_3$ theory 
was studied employing the method of dimensional regularization
\cite{McK,HT,Huish,Hager}. This method simplifies the renormalization process.  
In particular, the calculation of divergences arising in the evaluation of
Feynman-graph integrals is greatly simplified. Therefore, 
it was possible to push the renormalization process to a
considerably high order \cite{McK,HT,Huish,Hager}. 
In dimensional regularization, Feynman-graph integral divergences appear, 
in three dimensions, 
as poles in $\ve=3-d$ (where $d$ is the space dimension). These divergences 
appear in cutoff regularization as terms involving $\log\L$ (where $\L$ is the 
wave-number cutoff or an equivalent parameter in real space \cite{Ivan,Kharuk}). 
The terms that involve powers of $\L$ in cutoff regularization 
do not appear at all in dimensional regularization, whether those terms are divergent 
or not. Thus, this regularization method is unsuitable for 
effective field theories, in which the cutoff is held finite. 
Moreover, the cutoff must be allowed to run in the exact renormalization group 
\cite{Wil-Kog}. 

Nevertheless, the calculation of integrals with dimensional regularization can be useful. 
This is because terms involving $\log\L$ play a special role in  
effective field theory and are 
related to {\em marginal} coupling constants, in particular. 
We shall see how terms involving $\log\L$ appear, in $(\lambda\phi^4+g\phi^6)_3$ theory, in 
the renormalization of coupling constants precisely linked to the 
dimensionless constant $g$, while the renormalization of $\l$ in 
$(\lambda\phi^4)_3$ theory involves no divergences 
(provided that one avoids IR divergences in the massless or scale-invariant limit)
\cite{Parisi,ZJ}. 

Of course, renormalization of mass $m$ (or correlation length) is necessary 
in $(\lambda\phi^4)_3$ theory to absorb positive powers of $\L$ 
(Sect.~\ref{2loop}). However, 
the renormalized mass $m$ (or the correlation length) can be 
considered 
as the basic dimensional parameter to 
which the coupling constants are referred, and it is thus in the renormalization of 
coupling constants where one can appreciate the degree of universality; that is to say, 
it is in the renormalization of coupling constants where the cutoff scheme 
dependence will materialize. 
Therefore, 
all cutoff scheme dependence will disappear in $(\lambda\phi^4)_3$ theory, 
unlike in 
$(\lambda\phi^4+g\phi^6)_3$ theory. 

To prevent any misunderstanding, let us caution that we do not intend 
to study here tricritical behavior and its RG fixed point, but the 
Wilson-Fisher fixed point corresponding to ordinary critical behavior, 
in spite of the fact that we consider the $(\lambda\phi^4+g\phi^6)_3$ theory. 
The first fixed point is related to a free (``trivial'') renormalized field theory, 
whereas the second is related to a fully interacting field theory. 
The role of $g$, in our case, is just to bring up the non-universal corrections that 
appear for $m \neq 0$. However, before we examine the role of $g$, 
let us briefly review some results of the ordinary $(\lambda\phi^4)_3$ theory for 
critical behavior.

\section{Results of the $(\lambda\phi^4)_3$ theory}
\label{phi4}

The field-theoretic approach to the calculation 
of the critical exponents of the three-dimensional Ising model 
based on the $(\lambda\phi^4)_3$ theory was developed quite early 
\cite{Baker,LeG-ZJ}. This approach, together with the epsilon expansion, 
are the usual methods of precision calculation of critical exponents and other universal 
quantities \cite{Parisi,ZJ}. As remarked above, 
the renormalization of $\l$ in $(\lambda\phi^4)_3$ theory involves no UV divergences. 
After defining the dimensionless coupling constant $u=\l/m$, the 
output of perturbation theory is the quotient of the bare and renormalized couplings 
$\l_0/\l$ as a power series of $u$ (this notation, though common, 
differs from the notations of Refs.~\onlinecite{Parisi,ZJ}). 
Thus, one obtains the function $u(\l_0/m)$ and, hence, 
the $\b$-function
\begin{equation}
\b(u) = m\left(\frac{\p u}{\p m}\right)_{\!\!\l_0}, 
\label{bfun}
\end{equation}
which can also be expressed as a power series of $u$. 
The equation for scale-invariance $\b(u) = 0$ has a trivial zero solution and a 
non-trivial solution $u^* \neq 0$, the latter being the starting point for the derivation of 
universal quantities. Note that both $\l_0/\l$ as a function of $u$ and the 
function $\b(u)$ are universal, since a UV cutoff $\L$ is not necessary 
to obtain them. 

The fact that the $\b$-function is derived from perturbation theory and is only 
known as a power series leads to complications in the computation of universal quantities, 
because the power series is not convergent but just asymptotic and 
$u^*$ is not sufficiently small. Even at the two-loop order, one already needs 
some manipulation to obtain sensible results \cite[\S 8.3]{Parisi}. 
Of course, the sophisticated five-loop computation of Baker et al \cite{Baker} 
required a resummation of the power series (which they carried out with the Pad\'e-Borel 
method). The use of sophisticated methods of summation of divergent 
power series is a crucial step in 
the perturbative calculation of universal quantities \cite[ch.~42]{ZJ}.
The six-loop $\b$-function 
yields $u^*=0.985$ \cite[\S 29.4]{ZJ}, 
the old value being $u^*=0.987$ \cite{Baker,LeG-ZJ} (note our different normalization 
of $\l$ and hence $u$).

Given that we are not concerned here with the limit that is strictly scale-invariant, namely, 
with the case $m=0$, it behooves us to examine the behavior of the 
perturbative power series for small but non-vanishing $m$, that is to say, 
for $u$ close to but smaller than $u^*$, such that $\b(u) \lesssim 0$.
The known six-loop $\b$-function \cite[\S 29.2]{ZJ} is derived from 
the six-loop perturbative result for $\l_0/\l$:
\begin{equation}
\frac{\l_0}{\l} = 1+\frac{9 u}{2 \pi }+\frac{575 u^2}{36 \pi ^2}+1.83421
   u^3+1.91785 u^4+2.16641 u^5+2.04432 u^6+O\left(u^7\right).
\label{l0-u}
\end{equation}
Obviously, for $u \simeq 1$, equation (\ref{l0-u}) provides hardly 
any information. 
In fact, while $u^*$ is given by $\b(u^*) = 0$, the quotient $\l_0/\l$ has to diverge as 
$m \ra 0$ and $u \ra u^*$ (see Ref.~\onlinecite[\S 8.1]{Parisi}, for details). This is also 
an aspect of universality, viewed as independence of actual values of bare couplings. 
For $u \lesssim u^*$, the quotient $\l_0/\l$ must be finite but it may still be 
not calculable through equation (\ref{l0-u}) (unless using a resummation).


At any rate, series (\ref{l0-u}) should be suitable for small values of $u$. For example, 
it is fine even for $u=0.5$, it being such that the asymptoticness property holds 
(with a small percentage error).  
However, we cannot quantify how far this value or any particular value are from criticality,
since $u=\l/m$ is only a quotient and the magnitude of $m$ is unknown. 
Nevertheless, the mapping between bare and renormalized parameters, that is,
the mapping $(m_0,\l_0) \ra (m,\l)$, is one to one (off criticality). 
Unfortunately, the full mapping involves the renormalization of mass, and we clash head on 
with the problem of non-universality, given that the renormalization of mass involves 
divergences when $\L\ra\infty$ \cite{Parisi,ZJ}. 
Notwithstanding, let us recall the effective field theory philosophy.

In the ERG, $\L$ is always finite and runs from an initial value,
say $\L_0$, to the final value $\L=0$. Along the way, the mass and coupling constants 
run from their bare to their renormalized values. 
Therefore, one can define $\b$-functions for these coupling constants, which 
refer to a flow with the cutoff $\L$, unlike the $\b$-function in (\ref{bfun}), 
which gives the effect that a change of $m$ has on $\l$, once renormalization 
has been carried out, for a given value of $\l_0$.  
Employing the ERG, 
one can carry out simple numerical experiments on how the $\L$-running proceeds, 
in particular, observing the 
dependence on cutoff scheme \cite{II}. Indeed, we can use the renormalized values 
obtained in Ref.~\onlinecite{II} to find out what magnitude of $m$ corresponds to 
some value of $u$ for some fixed $\l_0$. 

\subsection{Exact renormalization group calculations for fixed bare coupling}

In Ref.~\onlinecite{II}, the partial  differential equation of the exact renormalization group in the local potential approximation was integrated several times, for  
null initial coupling constants, except 
a fixed $\l_0$, and a variable $m_0^2$.  
Tuning the latter, we obtain suitable values of $m$, 
from values close to $m=0$ (criticality) to values far from it. Far from criticality, 
we have a small value of $u$, close to the RG Gaussian fixed point $u=0$.
Each set of integrations was repeated employing three forms of 
the ERG, corresponding to different cutoff schemes:   
the Wegner-Houghton sharp-cutoff equation \cite{Wegner-H} 
and two others devised for the ERG.  
The goal, naturally, was 
to discern the influence of non-perturbative regularization schemes on the accuracy of 
the results. Since the full effective action involves an infinite number 
of parameters, non-perturbative renormalization group calculations generally truncate the 
full set of coupling constants to a moderate number 
(8 constants were used in Ref.~\onlinecite{II}). 
This reduction implies that different regularization schemes yield different results, 
making the comparison useful.

Of course, one can compare the results of different regularization schemes 
against one another. Furthermore, it is useful to compare them to the perturbative 
value of the quotient $\l_0/\l$, which is independent of the regularization scheme. 
Eq.~(\ref{l0-u}) includes the effect of field renormalization and is not exact for this comparison, given that the non-perturbative renormalization group 
of Ref.~\onlinecite{II} was limited to the evolution of the effective potential, namely,  
to  the partial truncation named {\em local potential approximation}.
The two-loop expression that does not account for field renormalization and 
was used in Ref.~\onlinecite{II} is
\begin{equation}
\frac{\l_0}{\l} = 1+\frac{9 u}{2 \pi }+\frac{63 u^2}{4 \pi ^2} + O\left(u^3\right).
\label{l0-u2}
\end{equation}
Note that $575/36=15.97$ in Eq.~(\ref{l0-u}) has been replaced by $63/4=15.75$, 
introducing a difference smaller than 1.4\%. 
Moreover, the difference in $\l_0/\l$, with $u=0.5$, 
amounts to $1.4\% \times 0.5^2/(4 \pi ^2) = 0.035\%$, 
which is actually smaller than other uncertainties in the calculations \cite{II}. 

The ERG integrations of Ref.~\onlinecite{II}, 
with $g_0=0$, fixed initial $\l_0$, and variable final $m$, 
spanning a range of $m/\L_0$ between $0.01$ and $0.07$,  
gave similar results in the three different cutoff schemes. 
Let us note that $\L_0$, being a fixed scale in effective field theory, can be taken 
as the reference scale, like in Ref.~\onlinecite{II}, 
and we often assume $\L_0=1$ and write dimensional 
quantities as numbers (as already done by Wilson and Kogut \cite{Wil-Kog}, for example).
The value $m/\L_0=0.01$ roughly corresponds to a correlation length of 100 lattice sites in 
the Ising model, say. It seems indeed that it corresponds to the domain where
scale invariance does not exactly hold but there are many short-distance degrees of freedom
involved. Nevertheless, the corresponding value of $u$ in Ref.~\onlinecite{II} was 
$u=\l/m \simeq 0.4$ (the exact value depending somewhat on the cutoff scheme). 
This value is sensibly smaller than $u^* \simeq 1$ and, in fact, it is such that 
the asymptoticness property of Eq.~(\ref{l0-u}) approximately holds.

The lesson learnt is that it is meaningful to consider situations which are, 
on the one hand, close enough to criticality, but which are, on the other hand, 
such that they may involve sizeable non-universal corrections. In fact, 
corrections of the order of inverse powers of the cutoff were mentioned 
in Ref.~\onlinecite{II}, beginning with $m/\L_0$, although no concrete calculation was made, 
because such corrections were assumed to be relatively small, 
given the magnitude of $m/\L_0$. 
However, the error is not necessarily small when we 
consider $g_0 \neq 0$.



\section{Two-loop perturbative $(\lambda\phi^4+g\phi^6)_3$ theory} 
\label{2loop}

\setlength{\unitlength}{1cm}
\thicklines
\begin{picture}(10,4)
\put(0.5,2){\circle{2}}
\put(4.5,2){\circle{2}}
\put(5.91,2){\circle{2}}
\put(8.8,2){\line(1,0){1.4}}
\put(9.5,2){\circle{2.2}}
\end{picture}

The effective potential at the two-loop order can be calculated with 
the background-field method (as outlined in Ref.~\onlinecite{I}). 
In the case of the $(\lambda\phi^4)_3$ theory, which is super-renormalizable, 
the {\em superficial} UV divergences of Feynman graphs are limited to the two-loop order, 
namely, to the ``single bubble'', ``double bubble'' and ``sunset'' Feynman graphs 
(displayed above) \cite[ch 9]{ZJ}. They only need a mass renormalization. 
In contrast, the $(\lambda\phi^4+g\phi^6)_3$ theory is just renormalizable, so 
an infinite number of Feynman graphs have superficial UV divergences. 
Of course, at the two-loop order and with the background-field method, 
we only have the ``single bubble'', ``double bubble'' and ``sunset'' 
{\em vacuum} Feynman graphs, which yield:
\begin{align}
U_\mathrm{eff}(\f) = U_\mathrm{clas}(\f)
+ \frac{h}{2}\, I_1
+\frac{3\, h^2\, U_\mathrm{clas}^{(4)}(\f)}{4!}\, I_2^2
-\frac{3 \,h^2 \,U_\mathrm{clas}^{(3)}(\f)^2 }{(3!)^2}\,I_3 + O\!\left(h^3\right),
\label{Ueff2loop1}
\end{align}
with the integrals
\begin{align}
I_{1}\!\left(M^{2}\right)=\int\frac{d^{3}{k}}{\left(2\pi\right)^{3}}\ln\left({k}^{2}+M^{2}\right),
\label{I1}\\
I_{2}\!\left(M^{2}\right)=\int\frac{d^{3}{k}}{\left(2\pi\right)^{3}\left({k}^{2}+M^{2}\right)}\,,
\label{I2}\\
I_{3}\!\left(M^{2}\right)=\int\frac{d^{3}{k}}{\left(2\pi\right)^{3}}
\frac{d^{3}{q}}{\left(2\pi\right)^{3}}\,
\frac{1}{\left({k}^{2}+M^{2}\right)\left({q}^{2}+M^{2}\right)\left({(\bm{k}+\bm{q})}^{2}+M^{2}\right)}\,,
\label{I3}
\end{align}
where $M^2 = U_\mathrm{clas}''(\f)$.
In Eq.~(\ref{Ueff2loop1}), $h$ denotes the loop-counting parameter (to be set to 1 at the end). 

We need to calculate 
$I_1$, $I_2$ and $I_3$. Actually, $I_2(M^2) = I_1'(M^2)$, 
so we need just two integrals.
They can be calculated with various cutoff schemes (see appendix \ref{integrals}). 
The calculation of the ``sunset'' integral $I_3$ is naturally more difficult. 
It has a logarithmic divergence, 
with a finite part which we denote by $A$. 
Suppressing inverse powers of $\L_0$, 
we have, in total, 
\begin{align}
U_\mathrm{eff}(\f) &= U_\mathrm{clas}(\f)
+ h 
\left(\frac{F_1 \L_0 U_\mathrm{clas}''(\f)}{4 \pi^2}-\frac{U_\mathrm{clas}''(\f)^{3/2}}{12 \pi}\right)
+\frac{3\, h^2\, U_\mathrm{clas}^{(4)}(\f)}{4!}
\left(\frac{F_1^2\L_0^2}{4 \pi ^4}-\frac{F_1\L_0 \sqrt{U_\mathrm{clas}''(\f)}}{4 \pi ^3}\right.
\nonumber\\
&+
\left.\frac{\left(8 B+\pi ^2\right)
   U_\mathrm{clas}''(\f)}{16 \pi ^4}\right)
-\frac{3 \,h^2 \,U_\mathrm{clas}^{(3)}(\f)^2 }{(3!)^2\left(16 \pi^2\right) }
\left(A-\log \frac{\sqrt{U_\mathrm{clas}''(\f)}}{\L_0}\right) + O\!\left(h^3\right).
\label{Ueff2loop}
\end{align}
With the sharp cutoff, 
we obtain $A=-1.5858$. 
We have also written the constant $B$ that arises in the finite part of $I_2^2$, and the 
constant $F_1$ that only affects $\L_0$ (see appendix \ref{integrals}). 
For the sharp cutoff, $B=1$ and $F_1=1$. 

We put
$$
U_\mathrm{clas}(\f) = \frac{m_0^2}{2}\,\f^2 + \l_0\,\f^4 + g_0\,\f^6
$$
(note that the this is not the usual normalization of coupling constants). 
By making $g_0=0$, we can compare with standard $(\lambda\phi^4)_3$ theory results 
(calculated at higher loop order and including the field renormalization factor $Z$) 
\cite{Soko-UO,Guida-ZJ}.

The renormalization is carried out as follows:
\begin{align}
m^2 &= U_\mathrm{eff}''(0) = m_0^2 
+h 
\left(\frac{6 \lambda _0 F_1 \Lambda _0}{\pi^2}
-\frac{3 m_0 \lambda _0}{\pi }\right) 
+ h^2 \left[
\frac{45 g_0 F_1^2 \Lambda _0^2}{2 \pi ^4}
-\frac{45 g_0 m_0 F_1 \Lambda _0}{2 \pi ^3}
\right.
\nonumber\\
&+
\left.
 \left(\frac{45}{8 \pi ^2} +\frac{45 B}{\pi^4}\right) g_0 m_0^2
-\frac{9  F_1 \Lambda _0 \lambda _0^2}{\pi ^3 m_0}
+\left(\frac{9 }{2 \pi ^2}+\frac{36 B}{\pi ^4}-\frac{6 A }{\pi ^2}\right)\lambda _0^2
+\frac{6 \lambda _0^2 }{\pi ^2}\log \frac{m_0}{\Lambda _0}
\right],
\label{mm02loop}
\end{align}
\begin{align}
\l = \frac{U_\mathrm{eff}^{(4)}(0)}{4!} &= \lambda _0 
+h \left(-\frac{15 g_0 m_0}{4 \pi }+\frac{15 g_0 F_1\Lambda _0}{2 \pi
   ^2}-\frac{9 \lambda _0^2}{2 \pi  m_0}\right) 
+h^2 \left[-\frac{315 F_1 \Lambda_0 g_0 \lambda _0}{4 \pi ^3 m_0}\right.
\nonumber\\
&+
\left.
\left(\frac{315 }{8 \pi ^2}+\frac{315 B}{\pi ^4}-\frac{30 A }{\pi ^2}\right) g_0 \lambda _0
+\frac{30 g_0 \lambda _0 }{\pi ^2}\log \frac{m_0}{\Lambda _0}
+\frac{27 \lambda _0^3 F_1\Lambda _0}{2 \pi ^3 m_0^3}
+\frac{18 \lambda _0^3}{\pi ^2 m_0^2}\right],
\label{ll02loop}
\end{align}
\begin{align}
g = \frac{U_\mathrm{eff}^{(6)}(0)}{6!} &= g_0 
+h \left(\frac{9 \lambda _0^3}{\pi 
   m_0^3}-\frac{45 g_0 \lambda _0}{2 \pi  m_0}\right) 
+ h^2 \left[
-\frac{675 g_0^2 F_1 \Lambda _0}{4 \pi ^3 m_0}
+ \left(\frac{675 }{8 \pi ^2} + \frac{675 B}{\pi ^4} - \frac{75 A }{\pi ^2}\right) g_0^2
\right.
\nonumber\\
&+
\left.
\frac{75 g_0^2 }{\pi ^2}\log \frac{m_0}{\Lambda _0}
+\frac{270 g_0 \lambda _0^2 F_1\Lambda _0}{\pi ^3 m_0^3}+\frac{225 g_0 \lambda _0^2}{\pi ^2 m_0^2}
-\frac{81 \lambda _0^4 F_1\Lambda _0}{\pi ^3 m_0^5}-\frac{108 \lambda _0^4}{\pi ^2 m_0^4}\right].
\label{gg02loop}
\end{align}
[The $O\!\left(h^3\right)$ symbol  
has been omitted, for brevity.] 

Solving for $m_0$ in Eq.~(\ref{mm02loop}) and substituting in the following two equations, 
we obtain:
\begin{align}
\l = 
\lambda _0 &- h\,\frac{3  \left(5 \pi  g_0 m^2-10 g_0 m F_1 \Lambda _0 + 
6 \pi  \lambda _0^2\right)}{4 \pi^2 m}
+ h^2 \left[
-\frac{135 F_1 \Lambda _0 g_0 \lambda _0 }{2 \pi ^3 m}
\right.
\nonumber\\
&+
\left.
\left(\frac{135 }{4 \pi ^2}+\frac{315 B}{\pi ^4}
-\frac{30 A }{\pi ^2}\right) g_0 \lambda _0
+\frac{30 g_0 \lambda _0 }{\pi^2} \log \frac{m}{\Lambda _0}
+\frac{99 \lambda _0^3}{4 \pi ^2 m^2}\right] + O\!\left(h^3\right),
\label{ll0r2loop}
\end{align}
\begin{align}
g =
g_0 &- h\,\frac{9  \left(5 g_0 m^2 \lambda _0-2 \lambda _0^3\right)}{2 \pi m^3}
+ h^2 \left[
-\frac{675 g_0^2 F_1\Lambda _0}{4 \pi ^3 m}
+ \left(\frac{675}{8 \pi ^2}+\frac{675 B}{\pi ^4}
-\frac{75 A}{\pi ^2}\right) g_0^2
\right.
\nonumber\\
&+
\left.
\frac{75 g_0^2 }{\pi ^2}\log\frac{m}{\Lambda _0} +
\frac{405 g_0
   \lambda _0^2 F_1\Lambda _0}{2 \pi ^3 m^3}+\frac{1035 g_0 \lambda _0^2}{4 \pi ^2 m^2}
-\frac{297 \lambda _0^4}{2 \pi ^2
   m^4}\right]+O\left(h^3\right).
\label{gg0r2loop}
\end{align}

Let us now solve for $\l_0$ in Eq.~(\ref{ll0r2loop}): 
\begin{align}
\l_0 &=
\lambda +
 h\frac{3\left(5 \pi  g_0 m^2-10 g_0 m F_1 \Lambda _0+6 \pi  \lambda ^2\right)}{4 \pi ^2 m}
\nonumber\\
&+h^2 \left[\left(\frac{30 A }{\pi^2} -\frac{315 B}{\pi ^4}
-\frac{30}{\pi ^2}\log \frac{m}{\Lambda _0}
\right) g_0 \lambda
+\frac{63 \lambda ^3}{4 \pi ^2 m^2}\right]+O\left(h^3\right).
\label{l0lr2loop}
\end{align}
This equation reduces to Eq.~(\ref{l0-u2}) when we take $g_0=0$. 
Such as it stands, it shows a dependence on $\L_0$, on the one hand, 
and on the constants $A$, $B$ and $F_1$, on the other hand, 
expressing non-universal corrections to Eq.~(\ref{l0-u2}) that 
appear for $g_0 \neq 0$.


From equations (\ref{gg0r2loop}) and (\ref{l0lr2loop}), 
we obtain $g(m,\l,g_0)$ as:
\begin{align}
g &= 
g_0-h\,\frac{9\left(5 g_0 m^2 \lambda -2 \lambda ^3\right)}{2 \pi m^3}
\nonumber\\
&+h^2 \left[\left(\frac{675 B }{\pi ^4}-\frac{75 A}{\pi ^2}
+\frac{75}{\pi ^2} \log \frac{m}{\Lambda _0}
\right) g_0^2
+\frac{1035 g_0 \lambda^2}{4 \pi ^2 m^2}
-\frac{27 \lambda ^4}{\pi ^2 m^4}\right]
+O\left(h^3\right).
\label{gg0l2loop}
\end{align}
This equation contains non-universal corrections, like Eq.~(\ref{l0lr2loop}), 
whenever $g_0 \neq 0$. For $g_0 = 0$, it simply becomes:
\begin{equation}
g = h\,\frac{9\lambda ^3}{\pi m^3} - h^2 \frac{27 \lambda ^4}{\pi ^2 m^4}
+O\left(h^3\right).
\label{gl2loop}
\end{equation}
One can compare it to Sokolov et al's results, \cite[eq. 2.5]{Soko-UO} or 
 \cite[eq. 8]{Soko-Kud}.

Of course, we can also obtain the function $g_0(m,\l,g)$ \cite{I}, which together with 
$m_0^2(m,\l,g)$ and $\l_0(m,\l,g)$ can be replaced in the expression (\ref{Ueff2loop})
of $U_\mathrm{eff}(\f)$ to yield a finite and $\L_0$-independent function in the limit 
$\L_0 \ra \infty$ (renormalizability). However, we will not need the 
renormalized expression of $U_\mathrm{eff}(\f)$.

\section{Calculation of non-universal corrections}
\label{non-univ}

The non-universal corrections that appear for $g_0 \neq 0$ can be evaluated through 
Eqs.~(\ref{l0lr2loop}) and (\ref{gg0l2loop}). First of all, let us reexamine how 
well the case $g_0=0$ reproduces the renormalization group integrations of 
Ref.~\onlinecite{II}, but making the comparison in a different manner. 
Since equation (\ref{l0lr2loop}), for $g_0=0$, reduces to Eq.~(\ref{l0-u2}), 
we first numerically solve this equation 
for $\l$, 
with given values of $m/\L_0$ and $\l_0/\L_0$. Then, we compute $g$  
by Eq.~(\ref{gl2loop}). 
The result is scheme independent. We should remark, of course, that scheme-dependent terms 
with negative powers of $\L_0$ have been suppressed in Eq.~(\ref{Ueff2loop}).  

It turns out that equation (\ref{l0-u2})
is not quite precise for $m/\L_0=0.01$, for example 
(the smallest value in Ref.~\onlinecite{II}). 
This is checked by considering the two next terms in Eq.~(\ref{l0-u2}), namely, 
the three and four-loop terms (ignoring the contribution of 
the field renormalization factor):
\begin{equation}
\frac{\l_0}{\l} = 1+\frac{9 u}{2 \pi }+ \frac{63 u^2}{4 \pi ^2} + 
{1.75448 \,u^3} + {1.75034 \,u^4} 
+ O\left(u^5\right).
\label{l0-u2_3l}
\end{equation}
Let us recall that $m/\L_0=0.01$ and $\l_0/\L_0=0.008225$ 
corresponded to $u=\l/m \simeq 0.4$ in Ref.~\onlinecite{II}.  
With $u=0.4$, we have
$$
1.75448\,u^3 = 0.1123,\quad 1.75034 \,u^4 = 0.04481;
$$
that is to say, non-negligible additions. We can neglect the $O\left(u^5\right)$ term. 

Note that the contribution of the field renormalization factor, that is to say, 
the use of Eq.~(\ref{l0-u}) instead of Eq.~(\ref{l0-u2}), would hardly alter these numbers. 
For example, $1.75448\,u^3 = 0.1123$ would have to be replaced by $1.83421\,u^3 = 0.117$, etc, 
with insignificant changes.

Let us first solve Eq.~(\ref{l0-u2}) for $\l$, 
with $\l_0=0.008225$ and $m=0.01$ (values from Ref.~\onlinecite{II}, normalized to $\L_0=1$). 
We obtain $\l=0.0043$ and hence $u=\l/m = 0.43$.  
This value is close to the results of 
the Wegner-Houghton ERG 
integration, $\l=0.0042$ and $\l/m = 0.42$ 
\cite{II}. 
Employing all the terms in Eq.~(\ref{l0-u2_3l}) and solving for $\l$, with the same values of 
$\l_0$ and $m$, we obtain $\l=0.0041$ and $u=\l/m = 0.41$. The difference with 
the WH ERG result is still a few percent, like only taking Eq.~(\ref{l0-u2}).



The magnitude of errors, i.e., few-percent errors, 
is compatible with the presence of terms of order 
$m/\L_0 \simeq 0.01$,   
which have been suppressed in Eq.~\ref{Ueff2loop} 
(such terms are scheme dependent).
At any rate, we expect that the dependence on $\L_0$ and on the regularization scheme 
will be greatly magnified when considering the sextic coupling.

\subsection{Renormalization with $g_0 \neq 0$ and sharp cutoff}
\label{g0neq0}

To examine the influence of having $g_0 \neq 0$ in the renormalization process, 
we need to solve Eqs.~(\ref{l0lr2loop}) and (\ref{gg0l2loop}) for $(\l,g)$, given 
a definite set $(m,\l_0,g_0)$. We can express those equations 
in terms of dimensionless variables $(u,g)$, but $m$ too appears now explicitly, unlike 
in the case $g_0 = 0$. 
We want $m/\L_0\simeq 0.01$ again, which gives, for example, $\log (m/\L_0)\simeq -4.6$ in 
Eqs.~(\ref{l0lr2loop}) and (\ref{gg0l2loop}). 
Of course, we must also set a regularization scheme that 
provides the constants $F_1$, $A$, and $B$ in Eqs.~(\ref{l0lr2loop}) and (\ref{gg0l2loop}). 

As done before with $g_0=0$, we employ the ERG to have an idea of the approach to 
criticality. This is done by first fixing $(m,\l_0,g_0)$, putting $(\l_0,g_0)$ 
in the ERG, and then tuning $m_0^2$ 
to reach the required renormalized $m$ in an ERG integration down to $\L=0$.
To be systematic, once fixed $(\l_0,g_0)$, 
we can find a suitable estimate of $m_0^2$ from an approximation of 
the WH equation, as shown below. 

Before proceeding, 
let us set scales for $\l$ and $g$ by considering their fixed-point values. 
Considering the 8th truncation of the Wegner-Houghton equation for the effective potential 
(up to $\phi^{16}$), as in Ref.~\onlinecite{II}, we have
$$\{m^2 = -0.4665,\, \l = 0.8189,\, g= 2.441,\, \ldots\}
$$
Here we are only interested in the coupling constants displayed (normalized to $\L_0=1$). 
Notice that the bare value of $m^2$ can be negative and 
this conventional notation is then improper, but we use it nonetheless. 
The exact fixed-point coordinates, without any truncation, have been obtained 
and are slightly different (e.g., $m^2 = -0.4615$)
\cite{Morris}.
The corresponding potential has two symmetric 
minima at $\f\simeq \pm 0.3$ and a maximum at $\f=0$. The parameters beyond $g$ hardly 
alter its shape for $|\f|<0.5$, while they produce a steeper growth for larger $|\f|$. 
It is naturally the field fluctuations between the two minima of the potential 
that flatten the maximum at $\f=0$ and 
give rise to a massless renormalized potential.

To have some clues about perturbative results 
we can refer to integrations of the WH RG equation.
%
We find it again convenient to begin with $(\l_0,g_0)$ close to the origin, 
like in Ref.~\onlinecite{II} (meaning much closer to the origin than the position of the 
ERG fixed point). 
In the space $(m_0^2,\l_0,g_0)$, the critical surface, where $m=0$, 
is well approximated near the origin by its tangent plane at the origin, 
which can be obtained from Eq.~(\ref{mm02loop}):
\begin{equation}
m_0^2 + h \frac{6 F_1 \Lambda _0}{\pi^2}\,\l_0
+ h^2 \frac{45 F_1^2 \Lambda _0^2}{2 \pi ^4}\, g_0 = 0.
\label{crit}
\end{equation}
Notice that the above-displayed fixed-point coordinates do not fit in this plane 
(after making $h=\L_0=F_1=1$). 
The reason is, of course, that the critical surface curves away from the origin.

To  work out a numerical example, 
we choose again $\l_0=0.008225$, like in Ref.~\onlinecite{II}, and we now set 
$g_0= (2\pi^2)^2\,0.00002= 0.007793$
(comparable to $\l_0$, the constant $A_3^{-1} = 2\pi^2$ coming from the WH equation). 
Equation~(\ref{crit}) gives $m_0^2 = -0.0068$ 
(with $h=\L_0=F_1=1$). 
By precisely tuning $m_0^2$, we find that a WH RG integration with 
$m_0^2=-0.00621$ yields 
$\l=0.005544$ and $m=0.0100834$ (which is small enough). 
Hence, $u=\l/m= 0.5498$.
The solution of Eq.~(\ref{l0lr2loop}) for $\l$, with the 
sharp-cutoff values of constants $A$, $B$, $F_1$, bare coupling constants 
$\l_0=0.008225$, $g_0= 0.007793$, and mass $m=0.01008$, yields $\l=0.005834$, 
that is to say, $u= 0.5785$. These values of $\l$ and $u$ agree with the WH RG result 
(within a few percent error).
However, they are quite different from the values obtained with the same 
$(m,\l_0)$ but with $g_0=0$.

We can also consider the influence of having $g_0 \neq 0$ on the 
renormalization of $g$ itself. 
Note that Eq.~(\ref{gl2loop}), 
in which $g_0=0$, gives $g \neq 0$ but only dependent on $u=\l/m$. In particular, 
taking the same $(m,\l_0)$ and $g_0=0$, we obtain $u=0.428$ and hence $g=0.13$. 
Taking $g_0= 0.007793$ instead, 
we obtain $g=0.29$. 
We can appreciate that a non-null $g_0$ greatly influences the renormalization of $g$, 
which is hardly universal for $m>0$. 
The WH RG integrations yield $g=0.27$ for $g_0= 0.007793$, and 
$g=0.14$ for $g_0= 0$. 
To explain why errors in $g$ are larger,  
we must consider that the perturbative series for $g$ in Eqs.~(\ref{gg0l2loop}) 
or (\ref{gl2loop}) are asymptotic and accurate only for very small 
$u$. This problem is quantified in Sect.~\ref{improved}. 

In fact, perturbation theory should not be expected to be reliable for $u>0.5$, 
in any event. 
In Sect.~\ref{improved}, we 
examine how reliable perturbation theory is for small $u$ and small $g_0$. 
We may guess that $g_0$ must be quite small; and indeed we have found that, for example, 
the value $g_0= (2\pi^2)^2\,0.0001= 0.03896$ is large enough already to ruin the 
agreement between the ERG and two-loop perturbation theory. 
Perhaps surprisingly, the non-universal terms in Eqs.~(\ref{l0lr2loop}) and 
(\ref{gg0l2loop}) are then considerable, especially, the two-loop contribution in 
Eq.~(\ref{gg0l2loop}). \Red{Let us see why.}

\Red{The relative magnitude  of the $g_0^2$ term in Eq.~(\ref{gg0l2loop}) is given by 
the coefficient (between parenthesis), which has a fixed part and a mass-dependent 
logarithmic part. Naturally, the latter dominates as $m \ra 0$, but let us assume 
that $m \lesssim \L_0$ and so $ |\log(m/\L_0)| \ll 1$. The fixed part, 
with sharp-cutoff constants, is $19.0$. Therefore, we should have $g_0 \ll 1/19.0 = 
0.053$.}
A deeper analysis is presented in Sect.~\ref{improved}, where 
two-loop perturbation theory is ``improved''.

As regards the renormalization of $\l$, which is our main focus, 
the main role of a $g_0 \neq 0$ is that 
it gives rise to a term proportional to $\L_0$, which is a major contribution to 
Eq.~(\ref{l0lr2loop}). 
Naturally, the magnitude of that term is due to having $m/\L_0 \ll 1$. 
The term proportional to $\L_0$ has negative sign in Eq.~(\ref{l0lr2loop}) and 
makes a positive contribution to $\l$, making it considerably larger when $g_0>0$. 

The term proportional to $\L_0$ in Eq.~(\ref{l0lr2loop}) is multiplied by the constant $F_1$, 
which is equal to one for the sharp cutoff but could be reduced, presumably, 
by employing a convenient cutoff scheme. Could we actually reduce it? 
Let us try to find how. We shall consider the other scheme-dependent constants as well.

\subsection{Renormalization with $g_0 \neq 0$ in various schemes}
\label{cutoffs}

Let us focus on the term proportional to $\L_0$ in Eq.~(\ref{l0lr2loop}), 
as the major non-universal contribution, and on the corresponding constant $F_1$, namely,
$$
F_1=\int_0^\infty F(q)\,dq.
$$
Here, $F(q)$ is the ``form factor'' regularizing the propagator as
$$
\frac{F(k/\L_0)}{{k}^{2}+m^{2}}
$$
(the term ``form factor'' is borrowed from nuclear physics, where it defines 
the spatial extent of a charge distribution). 
We assume that $F(0)=1$ and $\lim_{x\ra\infty}F(x)=0$ \cite{Liao,Gold},
and we further assume that $F(q)$ is non-increasing and $F'(0)=0$ 
(regarding the definition of the constant $F_2$, see appendix \ref{integrals}). 
These are natural assumptions to prevent pathological behavior and to follow the 
original idea of a smooth integration of Fourier modes \cite{Wil-Kog}. 
Furthermore, we assume that the derivative of the form factor function has 
a simple shape, as a regularized delta function of limited width 
that implements Wilson's idea of Fourier-mode shell integration.

In particular, we shall consider three types of functions $F(\ve,q)$ that depend on 
a parameter $\ve$ (the width) 
and that converge towards the sharp-cutoff function when $\ve \ra 0$; namely, 
we shall consider Liao et al's cutoff functions \cite{Liao}. In each case, $F(\ve,q)$ 
fulfills $F(\ve,1)=1/2$, for any $\ve$, so that the cutoff $\L_0$ marks a clear separation 
between ``almost completely integrated'' modes and ``not terribly integrated'' modes, 
in Wilson and Kogut's words \cite{Wil-Kog}. 

\subsubsection{Hyperbolic tangent}

$F(\ve,q)$ is a hyperbolic tangent modified to fulfill the above conditions:
$$
F(\ve,q) = \frac{1}{2} \left(\tanh\left(\frac{q^{-1}-q}{\ve}\right)+1\right).
$$
Any $\ve>0$ is possible, in principle, but it is sensible to restrict it, for example, 
to $\ve\leq 1$. 
Numerical computations of $F_1$ show that it is larger than one but close to one 
for $\ve=1$ and tends to one as $\ve \ra 0$, never being smaller than one. 

\begin{figure}
\includegraphics[width=8cm]{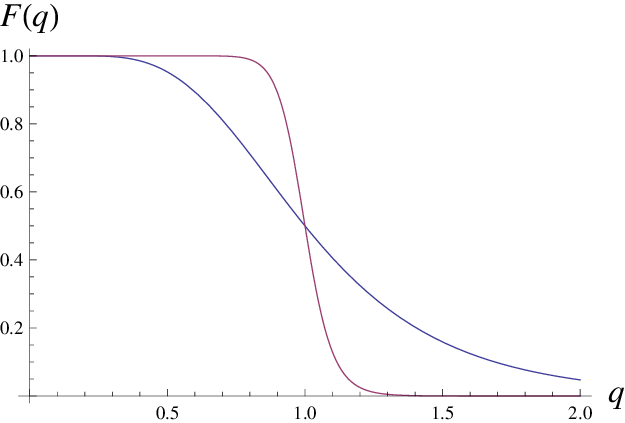}
\includegraphics[width=8cm]{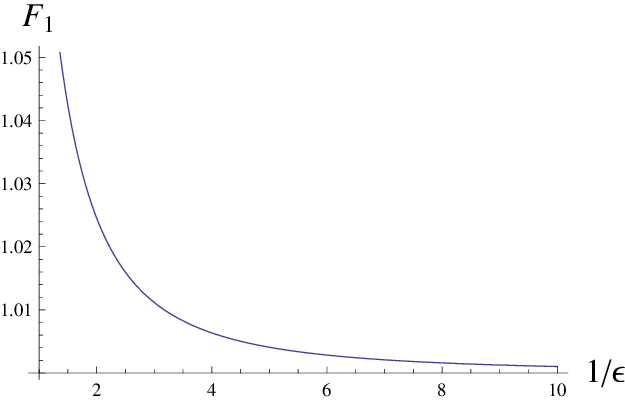}
\includegraphics[width=8cm]{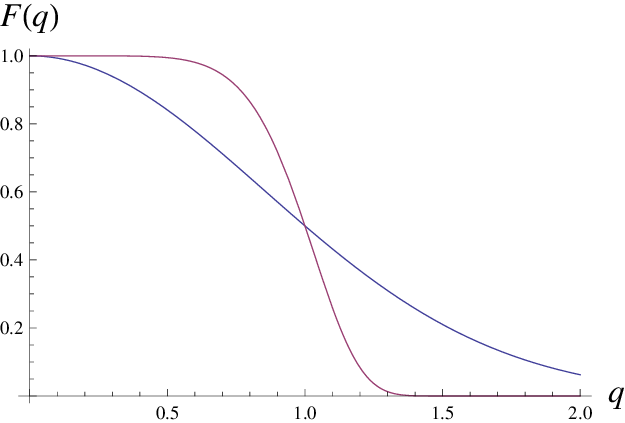}
\includegraphics[width=8cm]{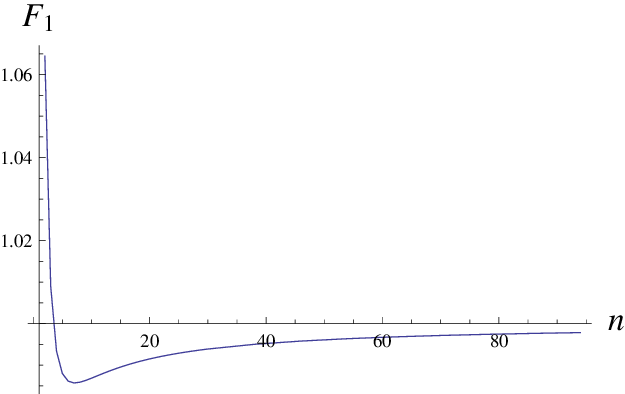}
\includegraphics[width=8cm]{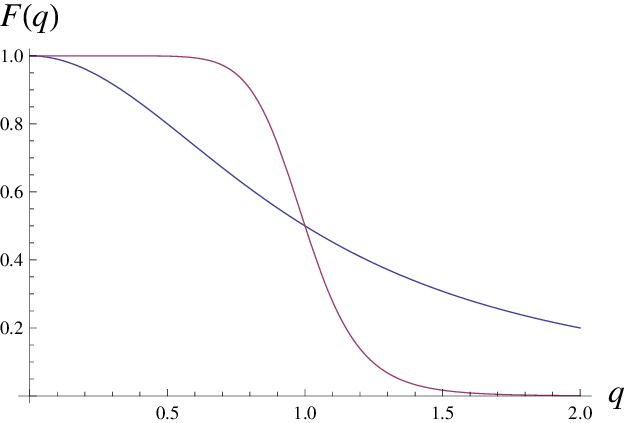}
\includegraphics[width=8cm]{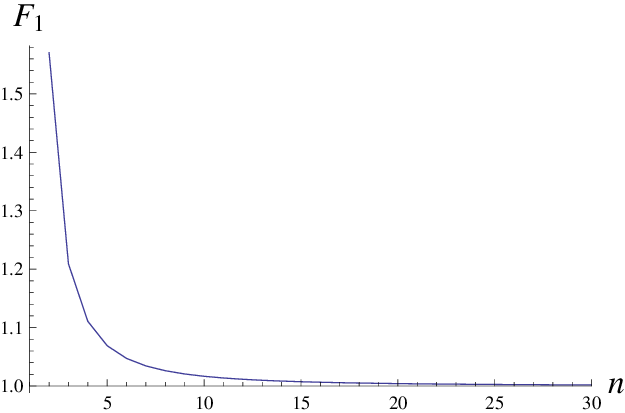}
\caption{The three types of cutoff functions and the corresponding values of $F_1$ plotted against cutoff function widths. First row: hyperbolic tangent  
(plotted for $\ve=1,1/5$); second row: exponential (plotted for $n=2,7$); 
third row: power law (plotted for $n=2,10$).}
\label{fig0}
\end{figure}

\subsubsection{Exponential function}

$F(\ve,q)$ is given by the exponential function:
$$
F(n,q) = 2^{-q^n},
$$
where the natural number $n\geq 2$ substitutes for the real variable $\ve$, and the limit 
$n \ra \infty$ substitutes for the limit $\ve \ra 0$. Again, 
$F_1$ is larger than one but close to one for $n=2$, 
\Red{namely, $F_1=\frac{1}{2} \sqrt{{\pi }/{\log (2)}}=1.064$,
and $F_1$ tends to one as $n \ra \infty$.  
However, $F_1$ is not monotone with $n$ and, for $n=7$, has the minimum 
$F_1={\Gamma \left(\frac{8}{7}\right)}/{\log ^{{1}/{7}}2}=0.9857$ (Fig.~\ref{fig0}). 
This very small reduction is offset by the complications of calculating $I_3$ and other 
Feynman integrals with that function.}

\subsubsection{Power-law function}

$F(\ve,q)$ is based on a power-law function:
$$
F(n,q) = \frac{1}{1 + q^n},
$$
where again the natural number $n\geq 2$ substitutes for the real variable $\ve$.
The dependence of $F_1$ on $n$ also presents a decreasing trend, never being smaller than one.
It takes the value $\pi/2$ for $n=2$, which is not very close to one, but it rapidly 
tends to one. Furthermore, for $n=2$, $F_2=\pi/2$ as well.

The power-law cutoff function is interesting for several reasons. It lends itself to 
analytical calculations, allowing for simple calculations of the other scheme-dependent 
constants $F_2$ and $A$ (see appendix \ref{integrals}). 
It is also useful for leading to simple forms of the ERG if 
$n=2,4$ (in dimension $d=3$) \cite{Liao}. The ERG for $n=4$ has been tested 
in Ref.~\onlinecite{II}, under the name 
of Morris's scheme. However, we are now more interested in the case $n=2$, as a very smooth 
cutoff that departs the most from the sharp cutoff.

\subsubsection{Conclusion of the examination}

\Red{A graphical summary is displayed in Fig.~\ref{fig0}.} 
Among Liao et al's cutoff functions, we cannot find any one that \Red{substantially} 
reduces $F_1$ from its sharp-cutoff value. 
Of course, this does not mean at all that such functions do not exist, 
because it is easy to find them, although they do not look natural, in the sense of 
Wilson and Kogut \cite{Wil-Kog}. A simple option is a function that 
almost completely integrates UV modes and, furthermore, integrates a good deal 
of modes with $k<\L_0$ (which may thus be ``terribly integrated''). 
At any rate, $F_1>1/2$, because $F(q)$ is non-increasing and $F(1)=1/2$. 
In fact, the derivative of a function like that must have an unphysical shape, 
that is, it must have a ``bump'' near $q=0$, in addition to the ``bump'' near $q=1$ 
that corresponds to the $k$-shell integration.

\section{Improved perturbation theory}
\label{improved}

The perturbative renormalization group is traditionally employed as 
a convenient method of improving the results of plain perturbation theory. For example, 
in $(\lambda\phi^4)_3$ theory, the beta function (\ref{bfun}) at the one-loop order is given 
by the first non-trivial term of series (\ref{l0-u})  
\begin{equation}
\b(u) = -u + \frac{9}{2\pi}\,u^2,
\label{b-fun}
\end{equation}
which is simply integrated as 
$$
\frac{u}{1-9u/(2\pi)} = \frac{\k}{m}
$$
(with an arbitrary constant $\k$). Thus, we can write the following expression 
for the RG transformation between an initial $m_0$ and an arbitrary $m$: 
\begin{equation}
\l_0 = \frac{\l}{1-(1/m-1/m_0)\,9\l/(2\pi)}\,. 
\label{b-fun-flow}
\end{equation}
Let us caution that $m_0$ is here some value of the {\em renormalized} mass, 
unlike the {\em bare} mass $m_0$ of preceding sections, which  
was such that $m_0^2<0$ in our examples. Likewise, 
$\l_0$ here is the renormalized value of $\l$ corresponding to $m_0$, 
not the bare value of $\l$. 
We use a common notation because both the ERG and the perturbative RG describe the 
change of parameters with scale, the scale being $\L$ in the former while being 
$m$ in the latter. In either case, the naught-subscript refers to initial conditions. 
Leaving aside the flow of $m$ in the ERG, the respective flows of $\l$ in each RG 
are related and a relationship between bare values and initial renormalized values can be 
established, as explained in the following.

Let us consider the flow towards masslessness, starting from some $u_0=\l_0/m_0 \ll 1$, 
such that $\b(u_0) \simeq -u_0$. While $m$ shrinks, $u$ grows but 
$\l$ hardly changes, until $m$ approaches $\l_0$ (from above). 
Afterwards, it undergoes a crossover and, 
for $m \ll \l_0$, Eq.~(\ref{b-fun-flow}) is ruled by the non-trivial fixed point, becomes independent of $m_0$, and can be expanded as
\begin{equation}
\frac{\l_0}{\l} = 1 + \frac{9\, u}{2\pi}+ \frac{81\,u^2}{4\pi^2} + \cdots
\label{bubble}
\end{equation}
This expansion has two remarkable aspects. Since it gives the function $\l(\l_0,m)$, 
independent of $m_0$, it constitutes a renormalization formula, with $\l_0 \gg \l$ 
playing the role of ``bare'' coupling. 
Naturally, 
$$\lim_{m\ra\infty}\l(\l_0,m)=\l_0\,.$$ 
In addition, the expansion is a ``guess'', namely, an RG improvement of the one-loop 
perturbative result that consists in the first non-trivial term of series (\ref{l0-u}). 
The second non-trivial term is guessed as $2.05u^2$, while it actually is $1.62u^2$ 
in Eq.~(\ref{l0-u}). 
In fact, the geometric series (\ref{bubble}) is the 
{\em bubble approximation}, which sums all the multi-loop Feynman graphs 
derived from the one-loop graph.

As a more sophisticated example, let us integrate the two-loop beta function derived from 
series (\ref{l0-u}). We now obtain:
$$
\frac{\l_0}{\l} = 1+\frac{9 u}{2 \pi }+\frac{575 u^2}{36 \pi ^2}+\frac{1109 u^3}{24 \pi ^3}
+\frac{22777 u^4}{216 \pi^4}+O\left(u^5\right).
$$
The $O\left(u^2\right)$ term agrees with Eq.~(\ref{l0-u}), of course, 
and the next term is guessed 
as $1.49u^3$, being $1.83u^3$ in Eq.~(\ref{l0-u}). The ``guessed'' terms actually correspond 
to a set of multi-loop Feynman graphs derived from two-loop graphs. 

\Red{Let us remark that the partial summation of Feynman graphs entailed by the integration 
of a few initial terms of the beta function is unrelated to Pad\'e-Borel 
summations that are performed on the beta function itself and change its properties,   especially, they change the position of the fixed point $u^*$.}

We can as well employ the two-loop perturbative $(\lambda\phi^4+g\phi^6)_3$ theory results 
in Sect.~\ref{2loop} for a similar improving procedure, 
which we suppose that should be in better accord with the results of the ERG. 
However, the procedure is not so simple.

\subsection{Improving $(\lambda\phi^4+g\phi^6)_3$ perturbation theory}

We need the two-loop perturbative beta functions for dimensionless coupling constants 
$u=\l/m$ and $g$, which are obtained from Eqs.~(\ref{ll0r2loop}) and (\ref{gg0r2loop}) in 
appendix \ref{betas} and read:
\begin{align}
\b_1 &= m\left(\frac{\p u}{\p m}\right)_{\!\!\l_0,\,g_0} = -u 
-\frac{15 g}{4 \pi}+ \frac{9 u^2}{2 \pi}
-\frac{165 g u}{8 \pi^2} + \frac{99 u^3}{4 \pi ^2}\,,
\label{b1}
\\
\b_2 &= m\left(\frac{\p g}{\p m}\right)_{\!\!\l_0,\,g_0} =
\frac{45 g u}{2 \pi}+ \frac{1275 g^2}{8 \pi ^2}
-\frac{27 u^3}{\pi}
-\frac{855 g u^2}{4 \pi ^2}+\frac{27 u^4}{\pi ^2}\,.
\label{b2}
\end{align}
We have set $h=1$ and ordered the monomials by their degrees. These differential 
equations are scheme-independent, as they have no trace of scheme-dependent constants. 

Before studying beta functions~(\ref{b1}) and (\ref{b2}), 
let us compare them with previous perturbative beta functions for 
$(\lambda\phi^4+g\phi^6)_3$ theory 
\cite{Law-Sar,Soko,McK,HT,Huish,Hager} (see also the beta functions for 
hypercubic models \cite[sect.~5]{Zin-Cod}).  
These beta functions have been usually employed only to analyze tricritical behavior, 
unlike what is being done here. 

Most previous beta functions do not take 
the renormalized mass but some other scale as RG parameter. In dimensional 
regularization, the RG parameter is the mass scale $\mu$ introduced to adjust the dimensions of renormalized quantities. 
In Lawrie and Sarbach's study of tricritical behavior 
\cite{Law-Sar}, the RG parameter $\l$ in Eqs.~(5.27--31) seems to be arbitrary, 
because they only say ``a change of length scale.'' 
However, they note that their beta functions are ``quite analogous'' to formulas obtained 
from the WH RG equations. 
However, the beta functions obtained from the WH RG equations are not polynomials in both 
the coupling constants {\em and} $m^2$, while those of Lawrie and Sarbach are. 
In fact, they add that they are ``restricted to studying tricritical singularities.''
In the same vein, the beta functions obtained by dimensional regularization 
are polynomials in the coupling constants and $m^2$ (which is understood 
as an additional coupling constant) \cite{McK,HT,Huish,Hager}. 
Our beta functions~(\ref{b1}) and (\ref{b2}) are instead polynomials in $u$ and $g$,  and therefore {\em not} polynomials in $m^2$.

At any rate, the main difference between our beta functions~(\ref{b1}) and (\ref{b2})
and the beta functions obtained by dimensional regularization is 
that the latter only contain contributions from the two-loop order onwards 
[they are $O(h^2)$] \cite{McK,HT,Huish,Hager}. 
As is well known, the one-loop terms are important for the Wilson-Fisher fixed point and,  
actually, give a fair approximation of $u^*$. In this regard, we will show below that 
the full two-loop beta function of $(\lambda\phi^4)_3$ theory, including the one-loop term,  
can be derived from Eqs.~(\ref{b1}) and (\ref{b2}). 

Our beta functions are similar to Sokolov's \cite{Soko}, 
which he proved to be suitable for describing the crossover between the tricritical and 
critical regimes. Sokolov's beta functions also have mass as RG parameter and are surely 
compatible with Eqs.~(\ref{b1}) and (\ref{b2}), once two differences are taken into account. 
First, Sokolov defines an ``auxiliary six-point diagram,'' which vanishes if $\l=0$, and 
hence subtracts pure-$\l$ contributions, thus actually redefining $g$. Second and more important, Sokolov truncates his beta functions to the second degree in 
the coupling constants, while ours have several terms of higher degree. 
Our calculation of the beta functions~(\ref{b1}) and (\ref{b2}) constitutes the 
first calculation of the {\em complete} two-loop beta functions of scalar field theory 
in three dimensions (to our knowledge).

The analysis of Eqs.~(\ref{b1}) and (\ref{b2}) begins by linearizing them, to 
have some information on the RG flow near the trivial fixed point. 
Since the linear $\b_2$ is null, $g=g_0$ (constant). Therefore, $\b_1$ gives
\begin{equation}
\l_0 = \l + \frac{15 g_0}{4 \pi} (m-m_0).
\label{lin-flow}
\end{equation}
One consequence of Eq.~(\ref{lin-flow}) is that initial values such that 
$g_0=-4\pi \l_0/(15 m_0)$ lead to $u = \l/m=\l_0/m_0=-15g_0/(4\pi)$. Naturally, the 
linear beta functions have a line of fixed points, due to the {\em marginal} coupling 
$(g\phi^6)_3$. 
If
\begin{equation}
0 < u_0+\frac{15 g_0}{4 \pi} \ll 1,
\label{FPline}
\end{equation}
then the flow 
is such that $g$ hardly changes while 
$$\l = -\frac{15 g_0}{4 \pi} \,m;
$$
that is to say, $\l$ decreases in absolute value linearly with $m$. Let us notice that 
an effective potential with negative $\l$ is bounded from below if $g>0$.

Equation~(\ref{lin-flow}) is to be compared with the linear part of 
Eq.~(\ref{l0lr2loop}), namely,
$$
\l_0 = \l + \frac{15 g_0}{4 \pi} \left(m-\frac{2 F_1\L_0}{\pi}\right).
$$ 
In this equation, 
the large non-universal term  corresponds to the $m_0$-dependent term in Eq.~(\ref{lin-flow}).
In fact, it seems natural to choose $m_0$ of the order 
of magnitude of $\L_0$, which is equivalent to having $F_1 \simeq 1$. 
Note that 
the limit $m_0 \ra \infty$ is not possible in Eq.~(\ref{lin-flow}), 
unlike in Eq.~(\ref{b-fun-flow}).
Thus, the presence of either $\L_0$ or $m_0$ is a signal of non-universality.

\begin{figure}
\includegraphics[width=10cm]{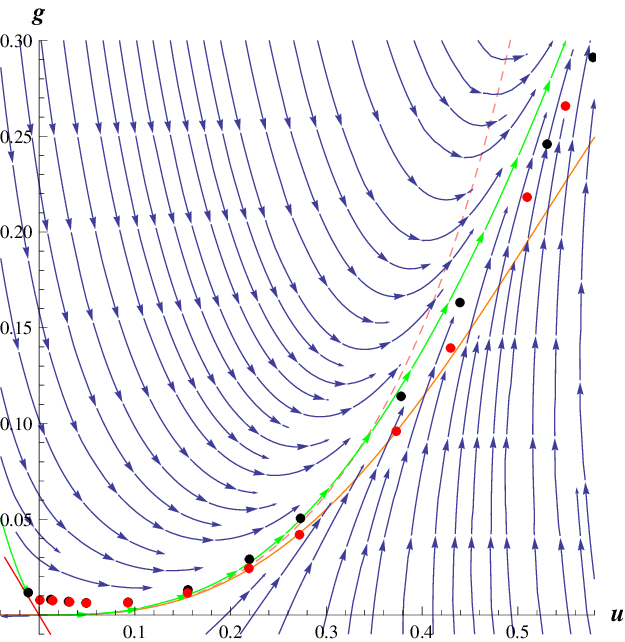}
\caption{RG flow for decreasing mass (in sharp-cutoff renormalization). 
Several features described in the text are displayed. 
Light red and black points correspond to Wegner-Houghton RG and two-loop calculations,  
respectively, nearly coinciding 
within some range 
(red points on top). The furthest couple of points has $m/\L_0=0.01$.}	
\label{fig1}
\end{figure}

Differential equations~(\ref{b1}) and (\ref{b2}) constitute an autonomous system when 
we take as independent variable the ``RG time'' $\tau = \log(m_0/m)$. They can be integrated 
numerically between $\tau=0$ and any $\tau>0$, from any point $(u_0,g_0)$. 
If we had the full beta functions rather than a low-order 
perturbative expression, then $u$ and $g$ should flow towards the 
non-trivial Wilson-Fisher fixed point. 
In contrast, what we observe is only that the flow converges, but the equations become 
inaccurate before approaching any non-trivial fixed point. 
Fig.~\ref{fig1} depicts the aspect of the flow.
The non-linear terms make the line of fixed points disappear, 
remaining the trivial fixed point $u=g=0$. In fact, the line of fixed points 
(Fig.~\ref{fig1}, short orange segment)
becomes a {\em separatrix} curve, in the terminology of dynamical systems
(short green line on the left).
This curve is actually the critical manifold of the trivial fixed point,
which corresponds to tricritical behavior.

The flow lines can also be obtained from the single differential equation 
$$\frac{dg}{du} = \frac{\b_2(u,g)}{\b_1(u,g)}\,,$$ 
with solution $g(u)$, given an initial condition, such as $g(0)=g_0$. 
Besides the separatrix, the other remarkable curve is given by 
the initial condition $g(0+)=0$ 
(specifying the $0+$ direction at the origin, as a singular point). It  
gives rise to the long green line with arrows in Fig.~\ref{fig1}. 
This is a ``connecting line'' in the terminology of dynamical systems. 

Our connecting line should be given by a function matching 
expansion (\ref{gl2loop}) (with $h=1$). The graph of expansion (\ref{gl2loop}) 
is the long orange line in Fig.~\ref{fig1}, below the long green line. 
It is easy to confirm that 
the derivative with respect to $u$ of expansion (\ref{gl2loop}) agrees with 
the expansion of $\b_2/\b_1$ in powers of $u$. 
However, the solution of the differential equation 
$dg/du = \b_2/\b_1$ is not a polynomial, of course, and it deviates from 
the polynomial (\ref{gl2loop}) as $u$ grows (as is visible in Fig.~\ref{fig1}). 
This non-polynomial solution $g(u)$ 
is supposed to constitute an {\em improvement} of the two-loop result. 
Assuming a power series for $g(u)$ and then matching coefficients, we find the 
next term in Eq.~(\ref{gl2loop}), namely, 
\begin{equation}
g(u) = \frac{9 u^3}{\pi }-\frac{27 u^4}{\pi ^2}+\frac{7209 u^5}{20 \pi ^3} 
+ O\left(u^6\right).
\label{gu2loop}
\end{equation}
The three-loop contribution computed by Sokolov et al.\  
is $(9/\pi)\times 1.39 = 3.98$ \cite[Eq 2.5]{Soko}, 
with the same sign as $7209/{(20 \pi ^3)}$ but quite different magnitude.  

Let us quote the known five-loop expression of the asymptotic expansion of $g(u)$ 
\cite[Eq 8]{Soko-Kud}:
\begin{equation*}
g(u) = \frac{9 u^3}{\pi }  \left(1-\frac{3 u}{\pi } +1.38996 u^2 -2.50173 u^3 
+ 5.2759 u^4 + O\left(u^5\right)\right).
\end{equation*}
Its graph is displayed in Fig.~\ref{fig1} as the dashed pink line. It matches the 
connecting line rather well 
up to $u=0.35$. 
In this regard, it 
should be noted that this five-loop expansion is not accurate at all 
in the entire range of $u$ in 
Fig.~\ref{fig1}: the two last terms in the parenthesis give, for $u=0.5$, respectively, 
$-2.50173\times 0.5^3 = -0.31$ and 
$5.2759\times 0.5^4 = 0.33$, with the same and considerable magnitude.

Other solutions of differential equations~(\ref{b1}) and (\ref{b2}) 
also constitute improved two-loop renormalization equations. 
In this sense, they improve, for example, on 
the pair of two-loop renormalization equations (\ref{ll0r2loop}) and (\ref{gg0r2loop})  
that have served to calculate Eqs.~(\ref{b1}) and (\ref{b2}). 
Equations (\ref{ll0r2loop}) and (\ref{gg0r2loop}) actually are the general 
solution of system (\ref{b1}, \ref{b2}), when this solution is restricted to 
$O\!\left(h^2\right)$, because they include two arbitrary 
integration constants, i.e., $\l_0$ and $g_0$. These integration constants are to be interpreted 
as bare coupling constants, and the solutions also involve the scheme-fixing constants. 
Thus, a change in the latter results in a change of the former, in order to have the same renormalized theory.

Universality is achieved on the connecting line, 
as independence of $m_0$, 
like we have seen before in $(\lambda\phi^4)_3$ theory. 
In other words, the limit $m \ra \infty$ ($\tau \ra -\infty$) is possible on it, 
because it leads to the trivial fixed point. This limit also leads to 
bare coupling constant values $\l_0 \neq 0$ and $g_0 = 0$, as in 
$(\lambda\phi^4)_3$ theory. In fact, 
the two-loop beta function of $(\lambda\phi^4)_3$ theory 
can be derived from Eqs.~(\ref{b1}) and (\ref{b2}), by substituting  
the relation $g(u)$ given by Eq.~(\ref{gu2loop}), namely, its first term (the other 
terms would contribute to higher order). 
To wit, by substituting the first term of Eq.~(\ref{gu2loop}) in Eq.~(\ref{b1}), 
we obtain
\begin{equation}
\b_1 = 
-u + \frac{9 u^2}{2 \pi} - \frac{9 u^3}{\pi ^2} + O\left(u^4\right).
\end{equation}
It is the correct beta function derived from Eq.~(\ref{l0-u2}) [Let us notice 
that the two-loop term in the beta function derived from Eq.~(\ref{l0-u}) instead is 
$-77 u^3/(9\pi ^2)$].


Even though there is no $m \ra \infty$ limit on any other RG trajectory, 
equations~(\ref{b1}) and (\ref{b2}) have one remarkable property: on the parabola 
\begin{equation}
g=6u^2/5, 
\label{cross}
\end{equation}
they reduce to 
$$\b_1 = -u, \quad \b_2 = 0.$$
As explained before, such form of $\b_1$ implies that $u$ changes with $m$ 
but $\l$ is stationary. 
The second equation obviously implies that $g$ is stationary.
Therefore, an RG trajectory, that is to say, a solution of Eqs.~(\ref{b1}) and (\ref{b2}), 
passing near the origin of the $(u,g)$-plane and crossing the curve $g=6u^2/5$ has 
stationary values of the coupling constants while 
near the crossing point (it spends there a considerable RG time $\tau$).
As there is no $m \ra \infty$ limit, if we let $m$ grow, then 
the RG trajectory eventually turns upward and 
$\l$ and $g$ change (the latter grows). Total stationarity only occurs on 
the connecting line.

The preceding considerations are useful to compare perturbation theory 
with the results of ERG integrations. 
In fact, 
the WH RG always gives $\l>0$ and hence $u>0$.
Furthermore, with bare $m_0>\L_0$, hardly any renormalization 
takes place in the integration over $\L$, and $(m,\l,g)$ keep their initial values. 
Therefore, letting $m$ grow, $u=\l/m$ shrinks along a horizontal segment in 
the $(u,g)$-plane towards the axis $u=0$ (Fig.~\ref{fig1}). 
In contrast, letting $m$ grow in perturbation theory makes $\l<0$ and hence $u<0$, 
while both $u$ and $g$ keep growing in absolute value. 

Referring to the bare values in the example of Sect.~\ref{g0neq0}, namely, 
$\{\l_0=0.008225,g_0=0.007793\}$, 
we find notable discrepancies between various WH RG integrations for $m_0>0$  
and the results of perturbative formulas 
(\ref{l0lr2loop}) and (\ref{gg0l2loop}) for the corresponding values of $m$. 
For example, for $m_0=3$, the WH RG integration yields 
$m=3.00003 \simeq m_0$, $\l=0.008430$, and $g=0.007791$, 
while Eqs.~(\ref{l0lr2loop}) and (\ref{gg0l2loop}) yield 
$\l=-0.01522$, and $g=0.00974$ (dimensional values in units of $\L_0$).
For $m_0=10$ (totally unphysical value), 
the ERG integration yields $m=m_0$, $\l=0.008244$, and $g=0.007793$,
while Eqs.~(\ref{l0lr2loop}) and (\ref{gg0l2loop}) yield 
$\l=-0.09067$, and $g=0.0105$. 
(See Fig.~\ref{fig1}.) 
At any rate, it is remarkable that major discrepancies only occur for 
unphysically large masses. For $m \lesssim 0.5$ 
the agreement is excellent: e.g., for $m_0=0.5^{1/2}=0.2326$, 
the ERG yields $m=0.2329$, $\l=0.01154$, and $g=0.006295$, while this $m$ and
Eqs.~(\ref{l0lr2loop}) and (\ref{gg0l2loop}) yield $\l=0.01145$, and $g=0.006347$.
%
The stationary point actually is $\{u=0.06601,g=0.005229\},$ and both 
$\l$ and $g$ hardly change in the interval $u \in (0.05,0.1)$, corresponding to 
$m \in (0.12,0.24)$. 

Of course, the agreement between the ERG and perturbation theory in the interval  
where $\l$ and $g$ hardly change (a horizontal segment in Fig.~\ref{fig1}) 
ceases for larger values of $u$ (smaller values of $m$), because 
the two-loop order is not sufficiently accurate. 
Nevertheless, the deviation is still 
relatively small even for $u=0.58$ ($m=0.01$, Sect.~\ref{g0neq0}), as 
seen in Fig.~\ref{fig1}.

The natural interpretation of all the above results is that the parabola (\ref{cross}) 
defines the crossover from tricritical to critical behavior. We can picture 
the parabola in Fig.~\ref{fig1} as the line formed 
by the minima of RG flow lines. The region between this line and the connecting line 
is the properly critical region whereas the region on the left is the tricritical region. 
Note that the exponent two of $u$ in Eq.~(\ref{cross}) is the 
classical value of the {\em crossover exponent}, as corresponds to the 
tricritical RG fixed point being a trivial fixed point
\cite{Law-Sar}.

\begin{figure}
\includegraphics[width=10cm]{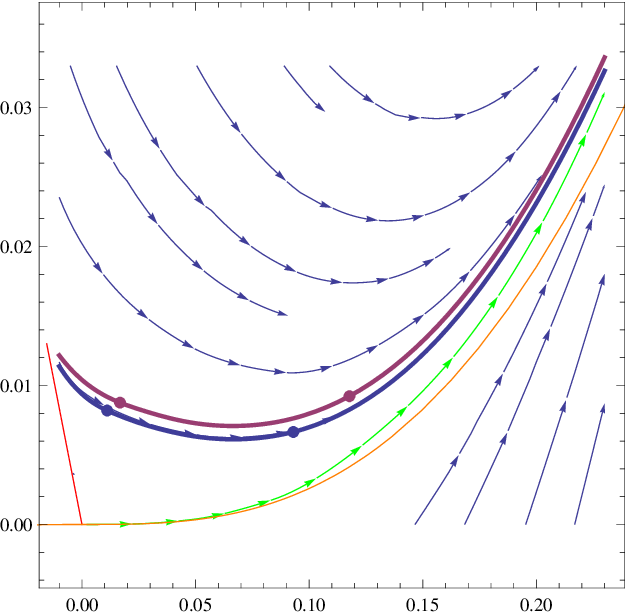}
\caption{Scheme dependence of perturbative results from 
Eqs.~(\ref{l0lr2loop}) and (\ref{gg0l2loop}) (interpolated). Sharp cutoff: lower (blue) line. 
Smooth $n=2$ power-law cutoff: upper (magenta) line. The two couples of points correspond to 
$m/\L_0=0.71$ and $m/\L_0=0.12$, respectively.}
\label{fig2}
\end{figure}

The scheme independence of the RG differential equations~(\ref{b1}) and (\ref{b2}) 
suggests that the example of Sect.~\ref{g0neq0}, namely, $\{\l_0=0.008225,g_0=0.007793\}$,  
when treated with a different cutoff, should yield similar results. 
Perturbative Eqs.~(\ref{l0lr2loop}) and (\ref{gg0l2loop}) contain the 
scheme dependence in the constants $F_1$, $A$ and $B$. The case of power-law cutoff 
with $n=2$ is interesting, as remarked in Sect.~\ref{cutoffs} 
(however, the results are similar in other cases). 
Since equations (\ref{ll0r2loop}) and (\ref{gg0r2loop}) are the general 
solution of system (\ref{b1}, \ref{b2}) to $O\!\left(h^2\right)$, 
keeping $\{\l_0,g_0\}$ while changing $F_1$, $A$ and $B$ should give rise to another 
solution. Indeed it does, and the solution is just slightly different, as shown in 
Fig.~\ref{fig2}. At any rate, the two trajectories are traversed somewhat differently, 
so that the same $m$ does not correspond to a pair of closest points.

Last, we can consider larger values of $g_0$, such as the one 
tried in Sect.~\ref{g0neq0}, namely, 
$g_0=0.039$ (five times larger than the first value). 
\Red{In this regard, we have shown that $g_0 \ll 0.05$ for the two-loop Eq.~(\ref{gg0l2loop}) 
to work. In addition, it is evident from Fig.~\ref{fig1} that the point with 
$\l_0=0.008225$ and $g_0=0.039$ 
is totally inside the tricritical region, as are points with very 
small $\l_0$ and $g_0$ approaching $0.05$. 
This restriction on the values of $g_0$ will surely not be altered by 
higher loop order calculations.}

\Red{Let us notice that the tricritical region is of interest, 
although we are concerned here with the crossover and critical regions. As mentioned above, 
we can consider effective potentials with $\l < 0$ and $g > 0$, which neatly lie in 
the tricritical region. These effective potentials can have three minima, as occurs 
on the {\em triple line} of the complete phase diagram \cite{Law-Sar}.}

\section{Conclusions}

The conclusions presented here are, of course, based on the two-loop effective potential, 
but they seem fairly general in perturbation theory. Certainly, further higher-loop order 
calculations will refine them. Nonetheless, we have confirmed our perturbative results by 
employing exact renormalization group integrations, within an approximation that 
is more accurate than two-loop perturbation theory.

The first fact is that three-dimensional scalar field theory is incomplete without 
the sextic field coupling and that a non-null bare coupling constant ($g_0$)
inevitably brings in non-universal results. While the Wilson-Fisher renormalization group 
fixed point has been thoroughly studied in perturbation theory with just 
the quartic field coupling, the introduction of the sextic field coupling is not 
only pertinent but also necessary. This is especially true 
in the effective field theory, 
because a null coupling constant $g_0$ at the cutoff scale $\L_0$ will 
not stay null at scales $\L<\L_0$, as ERG integrations demonstrate. 
Therefore, the question is the magnitude of the non-universal 
corrections due to having $g_0 \neq 0$.

The renormalization of the quartic and sextic coupling constants follows 
from the two-loop effective potential, calculated with an arbitrary cutoff function, 
which we denote by $F(k/\L)$ and also call the form factor. 
At the two-loop order, the scheme dependence, that is to say, 
the dependence on the cutoff function, reduces to 
three constants, which result from the Feynman graph integrals with  
the cutoff function (computed with variable difficulty). 
To wit, we define the constants 
$F_1$, $F_2$, and $A$, the last one being the most difficult (we also use $B = F_1 F_2$).
These constants appear in renormalization terms that explicitly contain 
the cutoff scale $\L_0$ in addition to the coupling constant $g_0$, 
and which do not have a finite limit when $\L_0 \ra \infty$.

The magnitude of scheme dependence is quantified by the allowed ranges of 
constants $F_1$, $F_2$, and $A$. In principle, these ranges could be large or even 
arbitrary, but a close examination of various types of cutoff functions shows 
that reasonable schemes only allow very reduced ranges. A remarkable result is 
that the smallest values of those constants occur precisely for the sharp cutoff. 
Thus, the non-universal terms are smallest in this case. However, smooth cutoffs 
do not raise much their magnitude. Suitable examples of very smooth cutoff 
functions are the ordinary exponential and the $n=2$ power-law function. We have 
chosen the latter, because it is more amenable to analytical treatment and 
would be suitable for simple exact RG integrations. 

Given the reduced ranges of scheme-dependent constants $F_1$, $F_2$, and $A$, 
the magnitude of the bare sextic coupling constant $g_0$ determines the magnitude of 
the non-universal terms. It is not obvious what magnitude of $g_0$ is appropriate for 
the renormalization formulas. Since we intended to begin with 
the sharp cutoff and compare with Wegner-Houghton RG integrations, 
we have used the criterion of small bare values, that is to say, 
values that allow an easy approximation of the critical manifold (and indeed of the 
Wegner-Houghton equation). 
This criterion has no fundamental reason, of course, but it makes it easier  
to find critical values of the coupling constants. 

Thus, we have initially tried $\l_0/\L_0$ and $g_0 < 0.01$. 
We do not attempt to determine exact critical values 
but only to reduce the renormalized mass to $m/\L_0 \simeq  0.01$. In this way, we find that 
$u = \l/m \gtrsim 0.5$, that is to say, values of $u$ that we expect to be still suitable for 
simple summation of perturbative series in $u$. 
We further find that the effect of a non-zero $g_0$ on 
both the renormalized coupling constants 
is considerable, and the effect is more pronounced on $g$ than on $\l$. 
Nevertheless, 
comparing two-loop calculations with exact RG integrations, we find that their concordance 
is not spoiled if $g_0 < 0.01$. In contrast, larger values of $g_0$ produce 
increasing discrepancies, which are very large for $g_0 > 0.03$ 
(still with $\l_0/\L_0 < 0.01$). 
The origin of these discrepancies becomes clear when we achieve a complete picture of 
how renormalization works for a wide mass range. 

To have a complete picture of the renormalization group, 
we have calculated the perturbative renormalization group that 
gives the changes of couplings with mass, at the two-loop order. 
This type of RG is interesting in itself, as it is, on the one hand, 
scheme independent, and, on the other, it allows for improvements in perturbation theory.
Naturally, the two-loop order loses accuracy far from the trivial RG fixed point. 
We have plotted its aspect in the region where $u \in (0, 0.6)$ 
and $g \in (0, 0.3)$ (and for very small negative $u$, in addition, see Fig.~\ref{fig1}). 
The RG flow plot clearly shows how the crossover from tricritical to critical behavior 
takes place. The condition that somehow determines the region 
of critical behavior, ruled by the Wilson-Fisher fixed point, is $g<6u^2/5$.  
Thus, trajectories like the one with $\l_0/\L_0 < 0.01$ and $g_0 > 0.03$ are mostly 
in the tricritical region. 

The results of renormalization for a definite couple of bare couplings, such that 
$\l_0/\L_0$ and $g_0 < 0.01$, and letting $m$ run from very large 
(larger than the cutoff $\L_0$) 
to quite small ($m/\L_0 \simeq  0.01$), can be best analyzed when set in the 
RG flow diagram. The small values of bare couplings guarantee that the renormalized 
coupling constants, for physical values 
of $m$, fall in the region of critical rather than tricritical behavior. However, the close 
concordance between two-loop calculations and exact RG integrations that occurs for 
small $\l_0/\L_0$ and $g_0$, in a mass range, say $m/\L_0 \in (0.1,0.5)$,  
is progressively lost for smaller values of $m$. 
Of course, one should expect to achieve, 
at criticality, full agreement between exact RG calculations and large-order perturbation theory, conveniently resummed. However, confirming this is beyond 
the reach of the approximations that we employ and is not our goal here.

We can also set the results of two-loop calculations for common $\l_0$ and $g_0$ but
different cutoffs in the RG flow diagram (Fig.~\ref{fig2}). 
Naturally, it is most interesting to contrast 
the sharp cutoff with a very smooth cutoff, such as the $n=2$ power-law cutoff. 
The renormalized trajectory that corresponds to the smooth cutoff is a little further away 
from the connecting line that represents the universal trajectory for $g_0=0$. 
The difference is small, as corresponds to the small range
of the scheme dependent constants. As the trajectories converge towards 
the Wilson-Fisher point, they become indistinguishable. 
However, points that are initially close, that is to say, 
close for $m/\L_0 \simeq 1$, become distant for $m \ll \L_0$. 
This property attributes a measurable meaning to the scheme dependence. 
Of course, this meaning relies on the knowledge of $\L_0$ as the reference scale. 


Since the main factor for non-universality is the magnitude of $g_0$ rather than the type of cutoff scheme, one might wonder whether $g_0$ should be small and why. 
Considering the usual methods 
of defining a field theory for the Ising model (Hubbard-Stratonovich transformation, etc), 
we see no reason why $g_0$ should turn out small. This suggests that 
the three-dimensional Ising model field theory could possibly lie in the tricritical region 
and would only enter the critical region for a sufficiently large correlation length. 

From a general point of view, the limited influence of the regularization scheme when 
the mass is sufficiently small can be useful in effective field theories for particle 
physics. The question is: given a fixed field theory ``beyond the Standard Model,'' 
how should the influence of regularization schemes on accessible low-energy physics be 
quantified?

\begin{acknowledgments}
I thank Gary Goldstein for various exchanges and for encouraging me 
to consider applying effective field theory to condensed matter problems. 
I am also grateful to Riccardo Guida and Andrey Kudlis for remarks on the renormalization 
of Phi$^4$ theory, and to Natalia Kharuk for comments on her preprint.
\end{acknowledgments}

\appendix
\section{Calculation of integrals}
\label{integrals}

Let us begin with $I_1$ and $I_2$. 
Since $I_2(M^2) = I_1'(M^2)$, we can compute one of them and hence deduce the other. 
In fact, 
$I_2$ is simpler. Of course, these integrals have to be integrated over the entire $k$-space  
but are UV divergent, that is to say, the integrals diverge for $k\ra\infty$. Let us 
regularize the field with a sort of ``form factor'', namely, a smoothing or  
average over a space volume with a characteristic length that is large enough to contain 
several lattice sites, say, but that is small enough with respect to the correlation length. 
Therefore, we can assume that its effect on the propagator is to multiply it by a 
factor $F(k/\L_0)$, such that $F(0)=1$ and $\lim_{x\ra\infty}F(x)=0$.  
Our integral becomes:
\begin{equation}
I_{2}\!\left(M^{2}\right)=
\int\frac{F(k/\L_0)\,d^{3}{k}}{\left(2\pi\right)^{3}\left({k}^{2}+M^{2}\right)}
= \frac{1}{2\pi^2} \int_0^\infty\frac{F(k/\L_0)\,k^2\,d{k}}{{k}^{2}+M^{2}}
\,.
\label{I2F}
\end{equation}
Naturally, the sharp cutoff is defined by $F(k/\L_0)=\th(1-k/\L_0)$, where 
$\th(x)$ is the Heaviside step function, but we shall keep $F$ as general as possible.

We can easily extract an $M$-independent term from $I_{2}$ in Eq.~(\ref{I2F}): 
\begin{equation}
I_{2}\!\left(M^{2}\right) - I_2(0) =
\frac{1}{2\pi^2} \int_0^\infty F(k/\L_0)\,k^2\,d{k}
\left(\frac{1}{{k}^{2}+M^{2}} - \frac{1}{k^2}\right) =
\frac{-M^2}{2\pi^2} \int_0^\infty \frac{F(k/\L_0)\,d{k}}{{k}^{2}+M^{2}}
\,.
\end{equation}
Therefore,
\begin{equation}
I_{2}(M^{2}) = \frac{1}{2\pi^2} 
\left(\int_0^\infty F(k/\L_0)\,dk - M^2\int_0^\infty\frac{F(k/\L_0)\,d{k}}{{k}^{2}+M^{2}}
\right).
\label{I2split}
\end{equation}
Note that we have achieved a useful separation: while the first integral diverges 
in the limit $\L_0\ra\infty$, the second one is finite and gives
\begin{equation}
\int_0^\infty\frac{{dk}}{{k}^{2}+M^{2}} = \frac{1}{M}\int_0^\infty\frac{{dq}}{{q}^{2}+1} = 
\frac{\pi}{2M}\,.
\label{I22}
\end{equation}
We also have
\begin{equation}
\int_0^\infty F(k/\L_0)\,dk = \L_0\int_0^\infty F(q)\,dq,
\end{equation}
where we can assume that the latter integral is finite and the UV divergence is 
proportional to $\L_0$. If $F(q)=\th(1-q)$, the integral is trivially equal to one.

Given that we have $I_2^2$ in Eq.~(\ref{Ueff2loop}), we need to calculate inverse 
powers of $\L_0$. To do it, we express the second integral
in Eq.~(\ref{I2split}) as 
\begin{equation}
\int_0^\infty\frac{F(k/\L_0)\,d{k}}{{k}^{2}+M^{2}} = 
\frac{1}{\L_0}\int_0^\infty\frac{F(q)\,d{q}}{{q}^{2}+(M/\L_0)^{2}}\,,
\label{I22M}
\end{equation}
and we try to expand the latter integral for small $M$. 
We already know that the integral diverges as $\L_0/M$ for small $M$, to fit 
Eq.~(\ref{I22}) 
The divergence of the integration over $q$ in (\ref{I22M}) for $M=0$ 
is an IR divergence at $q=0$. We can segregate it by expanding $F(q)$ 
at $q=0$. Let us assume that 
$$
F(q) = 1 + O\!\left(q^2\right),
$$
that is to say, that $F(q)$ is twice differentiable at $q=0$ and $F'(0)=0$ 
(a very natural assumption). 
Thus,
$$
\int_0^\infty\frac{F(q)\,d{q}}{{q}^{2}+(M/\L_0)^{2}} = \frac{\pi\L_0}{2M} +
\int_0^\infty\frac{F(q)-1}{{q}^{2}+(M/\L_0)^{2}}\,d{q},
$$
being the last integral finite for $M=0$.


Denoting the two integrals of $F$ by
$$
F_1=\int_0^\infty F(q)\,dq,\quad F_2=\int_0^\infty\frac{1-F(q)}{{q}^{2}}\,d{q},
$$
we can write
\begin{equation}
I_{2}(M^{2}) = \frac{1}{2\pi^2} \left(\L_0 F_1 - \frac{\pi M}{2} + F_2 \frac{M^2}{\L_0} + M\,o\!\left(M/\L_0\right)\right)
\end{equation}
(where the little-o symbol means slower growth).
Therefore,
\begin{align}
I_{1}(M^{2}) &= \int I_{2}(M^{2})\, dM^2 = \frac{1}{2\pi^2} \left(\L_0 F_1 M^2 - 
\frac{\pi M^3}{3} + M^3\,o\!\left(1\right)\right).
\\
I_{2}(M^{2})^2 &= \frac{1}{4\pi^4} \left(\L_0^2 F_1^2 - \pi F_1 M \L_0 +
\left(\frac{\pi^2}{4} + 2F_1F_2 \right) M^2 + M^2\,o\!\left(1\right)\right)\,.
\end{align}
(The unimportant integration constant has been suppressed.) Notice that 
the $\L_0$-independent term is not universal, as it contains the scheme-dependent constant 
$F_1F_2$.
We have already noticed that $F_1=1$ in the case of the sharp cutoff. It is also 
easy to see that also $F_2=1$ in this case.

\subsection{Calculation of $I_3$}

Integral $I_3$ is of the type corresponding to ``melon'' Feynman graphs, 
and it is called, in particular, either ``sunrise'' or ``sunset'' graph integral. 
Including the form factor, we have:
\begin{equation}
I_{3}\!\left(M^{2}\right)= \frac{1}{\left(2\pi\right)^{6}}
\int_0^\infty dk \int_0^\infty dq \int_0^{\pi} 
\frac{8\pi^2\,\sin\th\,d\th\;F(k)\,k^2\,F(q)\,q^2\,F\left(|\bm{k}+\bm{q}|\right)}
{\left({k}^{2}+M^{2}\right)\left({q}^{2}+M^{2}\right)
\left({k^2+q^2+2kq\cos\th}+M^{2}\right)}\,,
\end{equation}
where $\th$ is the angle between $\bm{k}$ and $\bm{q}$. This integral is awkward, 
because $F\left(|\bm{k}+\bm{q}|\right)$ depends on $\th$.
Therefore, it is not the best expression to calculate the cutoff integral, not even 
in the case of the sharp cutoff.

A method that combines position and Fourier space to extract ``melon'' graph divergences 
was proposed long ago \cite{Tuthill}. 
In fact, the unregulated integral is very simple in terms of the correlation 
function in position space,
\begin{equation}
I_{3}\!\left(M^{2}\right) =  \int d^3\!x \left[G(x)\right]^3. 
\label{I3G}
\end{equation}
Given the simple form of $G(x)$ in three dimensions, some results can be 
simply derived from this expression \cite[Appendix to Ch 5]{Parisi}.
Inserting the form factor, we replace $G(x)$ by $G_\L(x)$, which 
can be simplified by carrying out the angular integral:
\begin{equation}
G_\L(x) = \int \frac{d^3\!k}{(2\pi)^{3}} \frac{e^{i\bm{k}\cdot \bm{x}}}{k^2+M^2}\,F(k/\L) 
= \frac{1}{2\pi^2\,x}  \int_0^\infty dk \,\frac{k\,\sin(kx)}{k^2+M^2}\,F(k/\L).
\label{GL}
\end{equation}
Therefore,
\begin{align}
I_{3}\!\left(M^{2}\right) = \frac{4\pi}{(2\pi^2)^3}
\int_0^\infty \frac{dx}{x}
\left[\int_0^\infty dk \,\frac{k\,\sin(kx)}{k^2+M^2}\,F(k/\L_0)
\right]^3 
\nonumber\\
= \frac{1}{2\pi^5}
\int_0^\infty\! dp \,\frac{p\,F(p/\L_0)}{p^2+M^2}
\int_0^\infty\! dk \,\frac{k\,F(k/\L_0)}{k^2+M^2}
\int_0^\infty\! dq \,\frac{q\,F(q/\L_0)}{q^2+M^2}
\int_0^\infty \frac{dx}{x} \,\sin(px)\,\sin(kx)\,\sin(qx).
\end{align}
The last integral can be carried out in terms of Dirichlet integrals by 
expressing the product of sines as a sum of sines (of different arguments); namely,
$$
\sin(px)\,\sin(kx)\,\sin(qx)  
= \frac{1}{4} (\sin (x (-p+k+q))+\sin (x (p-k+q))+\sin (x (p+k-q))-\sin (x (p+k+q))).
$$
Hence,
\begin{align}
\int_0^\infty \frac{dx}{x} \,\sin(px)\,\sin(kx)\,\sin(qx) &=
\frac{\pi}{8}   (\text{sgn}(-p+k+q)+\text{sgn}(p-k+q)+\text{sgn}(k+p-q)-1)
\nonumber\\
&= \frac{\pi}{4}(\theta (-p+k+q)\,\theta (p-k+q)\,\theta (k+p-q)),
\end{align}
where $\text{sgn}$ and $\theta$ denote the sign and step functions, respectively,  
and the last equality holds in the positive octant.
Thus we obtain
\begin{equation}
I_{3}\!\left(M^{2}\right) = \frac{1}{8\pi^4}
\int_\text{region} dp\, dk\, dq \,\frac{p\,F(p/\L_0)}{p^2+M^2}
\,\frac{k\,F(k/\L_0)}{k^2+M^2}
\,\frac{q\,F(q/\L_0)}{q^2+M^2}\,,
\label{I3red}
\end{equation}
where the ``region'' is defined by $0\leq p\leq k+q$ and $0\leq k\leq p+q$ and 
$0\leq q\leq k+p$.

Equation (\ref{I3red}) is a convenient representation, suitable for obtaining 
the behavior in the limit $\L_0\ra\infty$. Naturally, the form factor tends to one in this limit and $I_3$ is UV divergent (e.g., as $p=k=q \ra \infty$). The divergent part, 
proportional to $\log(\L_0/M)$, can be readily obtained with standard methods 
\cite[Appendix to Ch 5]{Parisi}, yielding
$$
\frac{1}{16\pi^2}\log(\L_0/M).
$$
Here, we are interested in the finite part, after the subtraction of this term. 
To subtract it, we split the integral (\ref{I3red}) into a UV convergent part and an IR 
convergent part, by setting an arbitrary cutoff $\k$ such that $M\ll \k \ll \L_0$. 
Then, we make the wavenumber-scale integration explicit with 
the change of integration variables:
$$
p = P\,x,\;k = P\,y,\;q = P\,z,
$$
such that the integration measure becomes:
$$ dp\,dk\,dq = dx \,dy\, dz \,\d(x+y+z-1)\, P^{2}dP.
$$
Now we can take the limit $\L_0\ra\infty$ in the integral in $[0,\k]$
and the limit $M\ra 0$ in the integral in $[\k,\infty]$ and thus write:
\begin{align}
I_{3}\!\left(M^{2}\right) = & \frac{1}{8\pi^4}
\int_\text{region} dx\, dy\, dz\, \d(x+y+z-1)\,xyz \left[
\int_0^\k
\frac{P^5\,dP}{(P^2 x^2+M^2)(P^2 y^2 +M^2)(P^2 z^2 +M^2)}
\right.
\nonumber\\
&+ 
\left.
\int_\k^\infty \frac{dP}{P\,x^2\,y^2\,z^2}\,F(Px/\L_0)\,F(Py/\L_0)\,F(Pz/\L_0)
\right].
\label{I3split}
\end{align}
Of course, the first integral is still IR divergent when $M=0$
and the second integral is still UV divergent when $\L_0\ra\infty$. In fact, 
the first integral diverges as $\frac{\pi^2}{2}\log(\k/M)$
and the second as $\frac{\pi^2}{2}\log(\L_0/\k)$, yielding the 
term proportional to $\log(\L_0/M)$.

Let us now consider the two integrals in Eq.~(\ref{I3split}) in turn. 
The first one is independent of the form factor and can be computed exactly
for large $\k/M$. A somewhat cumbersome calculation yields:
\begin{align}
\int_\text{region} dx\, dy\, dz\, \d(x+y+z-1)\,xyz 
\int_0^\k \frac{P^5\,dP}{(P^2 x^2+M^2)(P^2 y^2 +M^2)(P^2 z^2 +M^2)} 
\nonumber\\
= \frac{\pi^2}{2}
\left(-\frac{21 \zeta (3)}{2\pi ^2} - \log 3 + {\log \left(\k/M\right)} \right) + 
O\left(\frac{M}{\k}\right)^2.
\end{align}

The second integral in Eq.~(\ref{I3split}) gives the dependence of 
$I_{3}$ on the UV cutoff function. The integral over $P$ can be calculated analytically 
for simple functions, e.g., for the Gaussian $F(k)=\exp(-k^2)$, but the 
integration over $\{x,y,z\}$ is best computed numerically. 
Gathering it all, for the Gaussian we have:
\begin{align}
I_{3} &= \frac{1}{16\pi^2}\left(-\frac{21 \zeta (3)}{2\pi ^2} - \log 3 
-\frac{\gamma }{2} + \frac{2}{\pi^2}\, 2.325132375178383 + \log \left(\L_0/M\right)\right)
\nonumber\\
&= \frac{1}{16\pi^2}\left(-2.1948849888150064 + \log \left(\L_0/M\right)\right).
\end{align}
Let us notice that our conditions on cutoff functions imply that 
$F(k/\L)=\exp(-k^2/\L^2)$ must be replaced by $F(k/\L)=\exp(-\ln\!2\;k^2/\L^2)$, which 
amounts to a redefinition of $\L$.

%

\subsubsection{Power-law form factors}

Let us consider
\begin{equation}
F(k) =  \frac{1}{1+k^n}\,, 
\label{plaw}
\end{equation}
for several values of $n$.
This form factor allows us to calcute the integral giving $G_\L$, Eq.~(\ref{GL}), 
in several cases, e.g., for $n=2$. Therefore, the use of Eq.~(\ref{I3G}) is 
preferable; namely,
$$
I_{3}\!\left(M^{2}\right) = \frac{1}{2\pi^5}
\int_0^\infty \frac{dx}{x}
\left[\int_0^\infty dk \,\frac{k\,\sin(kx)}{k^2+M^2}\,F(k/\L)\right]^3.
$$
In fact, to obtain its finite part, after the subtraction of $\log(\L/M)$, 
we can proceed as above, namely, 
we split the integral into a UV convergent part and an IR 
convergent part, by setting an arbitrary cutoff $R$, now in position space, 
such that $M\ll 1/R \ll \L_0$. 
Now we can make $M\ra 0$ in the integral in $[0,R]$ and 
take the limit $\L_0\ra\infty$ in the integral in $[R,\infty]$.

The latter is independent of the form factor and can be computed exactly, because 
$$
\lim_{\L\ra \infty} G_\L(x) = G(x) = \frac{\exp(-Mx)}{4\pi x}
$$
\cite[Appendix to Ch 5]{Parisi}. Therefore,
\begin{align}
\int_{x\geq R} d^3\!x \left[G(x)\right]^3 &=
\frac{1}{16\pi^2} \int_R^\infty \frac{dx}{x} \exp(-3Mx) 
= \frac{E_1(3MR)}{16\pi^2} 
\nonumber\\
&= \frac{1}{16\pi^2}\left[-\gamma - \log(3MR) + O(MR)\right].
\label{IRinfty}
\end{align}

To calculate the integral in $[0,R]$, we first need
\begin{equation}
\lim_{M\ra 0}\int_0^\infty dk \,\frac{k\,\sin(kx)}{k^2+M^2}\,F(k/\L) 
= \int_0^\infty \frac{dk}{k} \,\frac{\sin(kx)}{1+(k/\L)^n}\,.
\label{masslessP}
\end{equation}
For some values of $n$, the result can be found in 
Gradshteyn-Ryzhik's tables \cite[3.728--34]{GR} (or obtained by computer algebra). 
For example,
\begin{align}
\int_0^\infty \frac{dk}{k} \,\frac{\sin(kx)}{1+(k/\L)^2}
= \frac{\pi}{2}(1-\exp(-\L x))\,;
\\
\int_0^\infty \frac{dk}{k} \,\frac{\sin(kx)}{1+(k/\L)^4}
= 
\frac{\pi}{2} \left(1 - e^{-\frac{\L x}{\sqrt{2}}} \cos \left(\frac{\L x}{\sqrt{2}}\right)\right);
\\
\int_0^\infty \frac{dk}{k} \,\frac{\sin(kx)}{1+(k/\L)^6}
= 
\frac{\pi}{6}  \left(3-e^{-\L x}-2 e^{-\L x/2} \cos \left(\frac{\sqrt{3} \L x}{2}\right)\right).
\end{align}
In these cases, it is possible to evaluate 
$$
\int_{x\leq R} d^3\!x \left[G(x)\right]^3 = 
\frac{1}{2\pi^5}
\int_0^R \frac{dx}{x}
\left[
\int_0^\infty \frac{dk}{k} \,\frac{\sin(kx)}{1+(k/\L)^n}
\right]^3.
$$
However, the results are expressed as combinations of functions $E_1$ 
with several arguments multiples of $\L R$ plus a $\log(\L R)$ term. 
The expressions become cumbersome for $n>2$. 
In the large $\L$ limit, each integral diverges as $(\pi/2)^3\log(\L R)$, as expected. 

Therefore, we can obtain the finite part with a simple integral over $x \in [0,\infty]$; 
for example, with $n=2$:
$$
\int_0^\infty dx
\left[\frac{1}{x}
\left(1-e^{-x}\right)^3 -  \frac{1}{x+1}\right] = \gamma +\log 3-\log 8.
$$
Adding the finite part given by Eq.~(\ref{IRinfty})  
and combining the logarithms,
for $n = 2,4,6,$ we have:
\begin{align}
I_3 = \frac{1}{16\pi^2} \left(-\log 8 + \log \left(\L/M\right)\right) =
\frac{1}{16\pi^2} \left(-2.07944 + \log \left(\L/M\right)\right).
\\
I_3 = \frac{1}{16\pi^2} \left( \frac{3 \log 5}{8} - \frac{9 \log 2-3 \log 3}{4} + \log \left(\L/M\right)\right) =
\frac{1}{16\pi^2} \left(-1.78000 + \log \left(\L/M\right)\right).
\\
I_3 = \frac{1}{16\pi^2} \left( -\frac{7 \log 2 + 13 \log 3 - 2 \log 7}{9} + \log \left(\L/M\right)\right) =
\frac{1}{16\pi^2} \left(-1.69357 + \log \left(\L/M\right)\right).
\end{align}

\subsubsection{Sharp cutoff}

The sharp-cutoff function can be obtained as the limit when $n\ra\infty$ of $F$ in 
Eq.~(\ref{plaw}). Hence, we guess that the constant in $I_3$ to be added to 
$\log(\L/M)$ is the least negative of the $n$-sequence. We have in place of 
the integral in the rhs of Eq.~(\ref{masslessP}), a tabulated integral \cite[8.23]{GR}, 
namely, 
$$
\text{Si}(\L x) = \int_0^\infty \frac{dk}{k}\sin(kx) \,\th(1-k/\L).
$$
The finite part of the integral over $x \in [0,R]$ is obtained as before, namely:
$$
\int_0^\infty dx
\left[\left(\frac{2}{\pi}\right)^3\frac{\text{Si}(x)^3}{x}  
-  \frac{1}{x+1}\right].
$$
This integral is evaluated numerically and we have
$$
I_3 = \frac{1}{16\pi^2} \left(-1.58578 + \log \left(\L/M\right)\right).
$$

\section{Beta functions} 
\label{betas}

To calculate the beta-functions, the simplest procedure is as follows. 
Let us begin with Eqs.~(\ref{ll0r2loop}) and (\ref{gg0r2loop}), which give 
$\l$ and $g$ as functions of $\l_0$, $g_0$ and $m$. 
Now we simply take derivatives with respect to $m$, keeping fixed $\l_0$ and $g_0$.
\begin{align}
\left(\frac{\p \l}{\p m}\right)_{\!\!\l_0,\,g_0} &=
h \left(\frac{9 \lambda _0^2}{2 \pi  m^2}-\frac{15 g_0}{4 \pi
   }\right)+
h^2\left(\frac{135 {F_1} g_0 \lambda _0 \Lambda }{2 \pi ^3
   m^2}+\frac{30 g_0 \lambda _0}{\pi ^2 m}-\frac{99 \lambda
   _0^3}{2 \pi ^2 m^3}\right)+O\left(h^3\right),
\\
\left(\frac{\p g}{\p m}\right)_{\!\!\l_0,\,g_0} &=
h \left(\frac{45 g_0 \lambda _0}{2 \pi  m^2}-\frac{27 \lambda _0^3}{\pi 
   m^4}\right) 
\nonumber\\
&+ h^2\left(-\frac{1215 {F_1} g_0 \lambda _0^2 \Lambda }{2 \pi ^3
   m^4}+\frac{675 {F_1} g_0^2 \Lambda }{4 \pi ^3
   m^2}-\frac{1035 g_0 \lambda _0^2}{2 \pi ^2 m^3}+\frac{75
   g_0^2}{\pi ^2 m}+\frac{594 \lambda _0^4}{\pi ^2 m^5}\right) + O\left(h^3\right).
\end{align}

Now, we replace $\l_0$ and $g_0$, using Eq.~(\ref{l0lr2loop}) to $O\left(h\right)$
and
\begin{equation}
g_0 = g + h\,\frac{9 \left(5 g m^2 \lambda -2 \lambda ^3\right)}{2 \pi  m^3} 
+ O\left(h^2\right). \nonumber
\end{equation}
Thus, we obtain:
\begin{align}
\left(\frac{\p \l}{\p m}\right)_{\!\!\l_0,\,g_0} &=
h \left(\frac{9 \lambda^2}{2 \pi  m^2}-\frac{15 g}{4 \pi
   }\right)+
h^2\left(\frac{99 \lambda ^3}{4 \pi ^2 m^3}-\frac{165 g \lambda }{8 \pi
   ^2 m}\right)+O\left(h^3\right),
\\
\left(\frac{\p g}{\p m}\right)_{\!\!\l_0,\,g_0} &=
h \left(\frac{45 g \lambda }{2 \pi  m^2}-\frac{27 \lambda^3}{\pi 
   m^4}\right)
+ h^2\left(
\frac{1275 g^2}{8 \pi ^2 m}-\frac{855 g \lambda ^2}{4 \pi ^2
   m^3}+\frac{27 \lambda ^4}{\pi ^2 m^5}
\right) + O\left(h^3\right).
\end{align}
Hence, we obtain the beta functions of non-dimensional coupling constants 
$u=\l/m$ and $g$:
\begin{align}
\b_1 &= m\left(\frac{\p u}{\p m}\right)_{\!\!\l_0,\,g_0} =
\left(\frac{\p \l}{\p m}\right)_{\!\!\l_0,\,g_0}\!\! - \frac{\l}{m} =
\\
&= -u +
h \left(\frac{9 u^2}{2 \pi}-\frac{15 g}{4 \pi
   }\right)+
h^2\left(\frac{99 u^3}{4 \pi ^2}-\frac{165 g u}{8 \pi^2}\right)+O\left(h^3\right).
\\
\b_2 &= m\left(\frac{\p g}{\p m}\right)_{\!\!\l_0,\,g_0} =
\\
&= 
h \left(\frac{45 g u}{2 \pi}-\frac{27 u^3}{\pi}\right)
+ h^2\left(\frac{1275 g^2}{8 \pi ^2}-\frac{855 g u^2}{4 \pi ^2}+\frac{27 u^4}{\pi ^2}
\right) + O\left(h^3\right).
\end{align}

\end{document}